\documentclass[prb,superscriptaddress,twocolumn,amssymb,floatfix,amsmath,showpacs,pdftex]{revtex4}

\usepackage{amsfonts}

\usepackage{graphicx}
\usepackage{subfigure}
\usepackage{color}
\graphicspath{{./img/}{./}}

\usepackage{booktabs}

\usepackage{hyperref}

\newcommand{\td}{\mathrm{d}}
\newcommand{\te}{\mathrm{e}}
\newcommand{\ti}{\mathrm{i}}

\begin{document}
\title{Type-II Bose-Mott insulators}

\author{Aron J. Beekman}
\email{aron@riken.jp}
\affiliation{Instituut-Lorentz for Theoretical Physics, Universiteit Leiden, P.O. Box 9506, 2300 RA Leiden, The Netherlands}
\affiliation{Correlated Electron Research Group (CERG), RIKEN Advanced Science Institute, Wako, Saitama 351-0198, Japan}

\author{Jan Zaanen}
\affiliation{Instituut-Lorentz for Theoretical Physics, Universiteit Leiden, P.O. Box 9506, 2300 RA Leiden, The Netherlands}

\begin{abstract}
The Mott insulating state formed from bosons is ubiquitous in solid $^4$He, cold atom systems, Josephson junction networks and perhaps underdoped high-$T_\mathrm{c}$ superconductors. We predict that close to the quantum phase transition to the superconducting state the Mott insulator is not at all as featureless as is
commonly believed. In three dimensions there is a phase transition to a low temperature state where, under influence of an external current, a superconducting state consisting of a regular array of `wires' that each carry a quantized flux of supercurrent is realized.  This prediction of the 
``type-II Mott insulator'' follows from a field theoretical weak--strong duality, showing that this `current lattice' is the dual of the famous Abrikosov lattice of 
magnetic fluxes in normal superconductors. We argue that this can be exploited to investigate experimentally whether preformed Cooper pairs exist in high-$T_\mathrm{c}$ superconductors.
\end{abstract}


\maketitle

\section{Introduction}
The Yin--Yang mystique in Asian philosophy has found a remarkably literal incarnation in modern physics in the form of the duality principle \cite{Zee03}. An elementary example of the idea that `opposites form a unity' is the particle--wave duality of quantum mechanics. This was surpassed by the identification of the 
Kramers--Wannier or weak--strong duality structures in quantum field theory, eventually leading to the rich dualities of string theory \cite{Polchinski96}. A pedestrian
example that will play a role in the background of the present story is the electromagnetic duality, stating that in a world where magnetic monopoles have a
similar standing as electrical charges, the `opposite' electrical and magnetic universes are mathematical mirror images \cite{Dirac31}. 

There is yet a deeper level that 
becomes particularly explicit when dealing with the strongly-interacting quantum many-particle systems of condensed matter physics. Such systems will 
typically have ordered ground states breaking symmetry spontaneously, for instance the superconducting state. With mathematical topology
it is then  possible to identify field configurations that are uniquely associated with the restoration of the broken symmetry: the topological excitations, such as the 
Abrikosov vortices in a superconductor. Upon increasing the quantum fluctuations of the collective state (e.g. increasing charging energy in the superconductor) at some point 
the system will undergo a zero-temperature quantum phase transition (QPT) where the system `melts' into a quantum disordered state \cite{FisherLee89,Kleinert08,HerbutTesanovic96,CvetkovicZaanen06a,NguyenSudbo99,HoveSudbo00}. The weak--strong duality 
principle now prescribes that this quantum {\em disordered} state can always be viewed as some {\em ordered} state formed from the topological excitations
associated with destroying the order of the ordered state. 

The archetypical example is the ``vortex--boson'' or ``Abelian-Higgs'' duality in two space and one time dimensions (2+1D), associated with a system of interacting bosons living on a lattice, undergoing a superconductor--insulator QPT. The simplest microscopic model of relevance to this physics is the Bose-Hubbard Hamiltonian \cite{FisherEtAl89}, realized literally in cold atom systems for the neutral case \cite{GreinerEtAl02} and Josephson junction networks \cite{FazioVanderZant01,BruderFazioSchoen05} for electrically charged bosons. For neutral bosons it reads as
\begin{equation}
H_\mathrm{BH} = - t \sum_{\langle ij \rangle}( b^\dagger_i b_j +  b^\dagger_j b_i ) + U \sum_i n_i (n_i -1) -\mu \sum_i n_i,
\end{equation}
which can be straightforwardly extended to the charged case by coupling in the electromagnetic gauge fields. We specialize to the case with an integer number of bosons per lattice site (``zero chemical potential''). For small charging energy $U$ the bosons will condense into a superfluid/superconductor. However, when $U/t \approx 1$ a quantum phase transition occurs to a Bose-Mott insulator. The charging energy exceeds the kinetic energy with the effect that the  bosons localize: a Mott gap opens and the lowest lying excitations are the \emph{doublons} (extra boson) and \emph{holons} (missing boson). This Bose-Mott insulator is conventionally considered to be a completely featureless state, not breaking any symmetry.

However, the naive picture of localized bosons is flawed\footnote{A wave function of combined orthogonalized Wannier functions can never lead to a sign-free
ground state, as is required for a many-boson wave function. This was pointed out most recently by
P.W. Anderson in Ref. \onlinecite{Anderson11}.}, since the ground state energy will always be lowered by virtual exchange processes. Moreover, although often not realized, the true nature of the Mott insulator close to the QPT is revealed by the vortex duality perspective \cite{CvetkovicZaanen06a,Franz07,BeekmanSadriZaanen11,MrossSenthil11}. In 2+1D vortices are point particles, and they embody the virtual quantum fluctuations in the superconductor as closed loops of vortex--antivortex worldlines. These loops grow in size when approaching the QPT, to `blow out' at the transition and the Mott insulator corresponds with a tangle of free vortex and antivortex worldlines (Fig. \ref{fig:vortex duality}). Elegantly, the vortex--vortex interactions can be parametrized in terms of effective $U(1)$ gauge fields, and this tangle of worldlines is therefore identical to a relativistic (Higgs) superconductor, where the Higgs mass is coincident with the Mott gap, while the holon and doublon excitations have the same 
status as the `massive photons' of this dual vortex superconductor\cite{CvetkovicZaanen06a}. This suggests that there is more going on than the featureless state one infers from the strongly-coupled, atomic limit `canonical view'. 

However, one has to now consider the thermodynamics of the dual superconductor. The vortex superconductor is charged and therefore the interactions between the dual vortices in the dual condensate are short-ranged. Since these are particles in 2+1D, they will proliferate at any finite 
temperature. There is therefore  no thermodynamical phase associated with the dual superconductor: one recovers the featureless Mott insulator. 
This situation is drastically different in three space dimensions.   

\begin{figure*}
  \def\svgwidth{6cm}
 \setlength{\unitlength}{\svgwidth}
 \global\let\svgwidth\undefined

\subfigure[vortex worldline]{
 \begin{picture}(1,0.61784071)(0,0)%
    \put(0,0){\includegraphics[width=\unitlength]{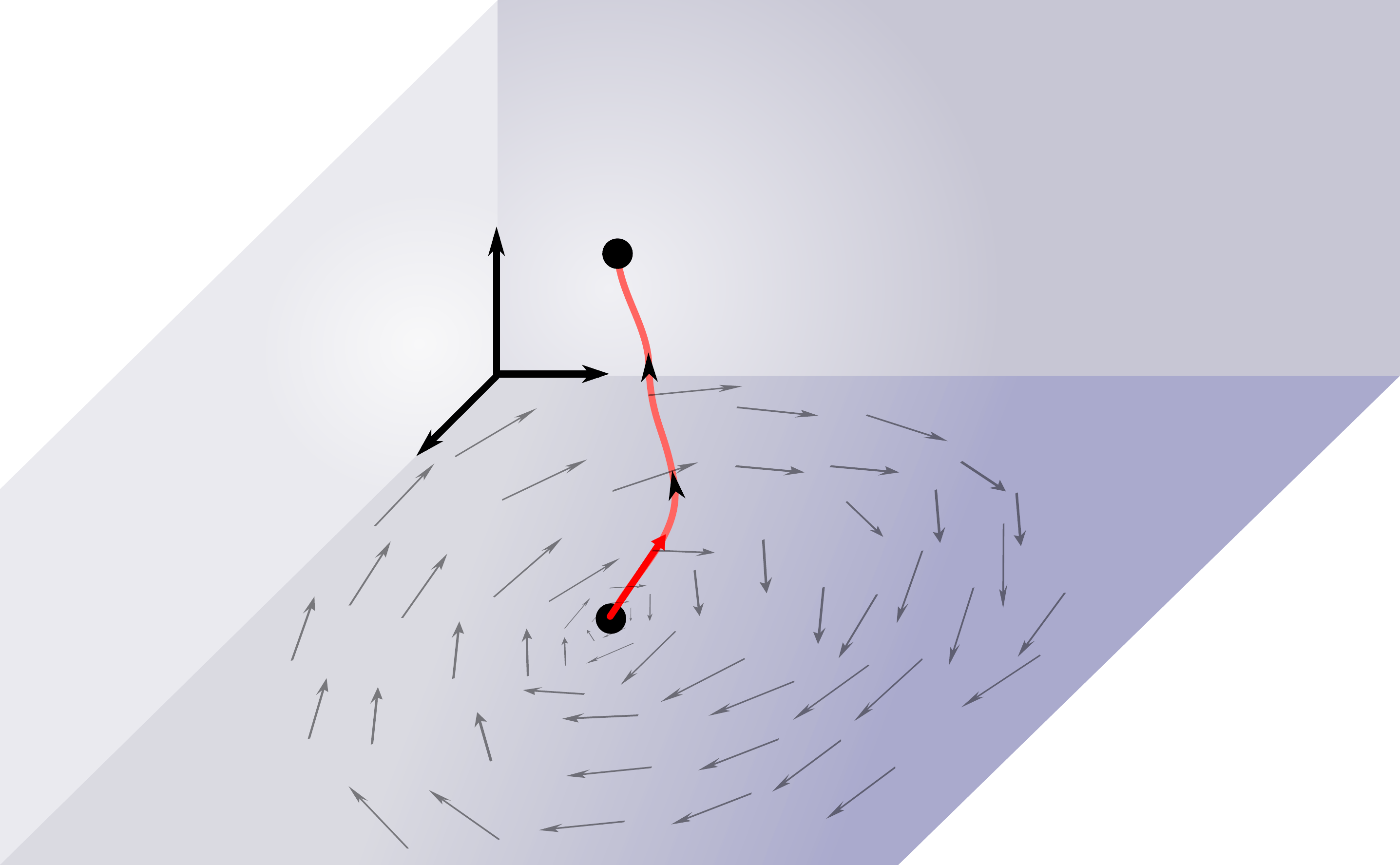}\label{fig:vortex worldline}}%
    \put(0.315,0.415){\color[rgb]{0,0,0}\makebox(0,0)[lb]{$t$}}%
    \put(0.39418673,0.36942472){\color[rgb]{0,0,0}\makebox(0,0)[lb]{$x$}}%
    \put(0.28,0.30){\color[rgb]{0,0,0}\makebox(0,0)[lb]{$y$}}%
    \put(0.48,0.19505591){\color[rgb]{0,0,0}\makebox(0,0)[lb]{$\kappa$}}%
  \end{picture}%
}
\hspace{1cm}
\subfigure[vortex worldsheet]{
  \begin{picture}(1,0.61800999)(0,0)%
    \put(0,0){\includegraphics[width=\unitlength]{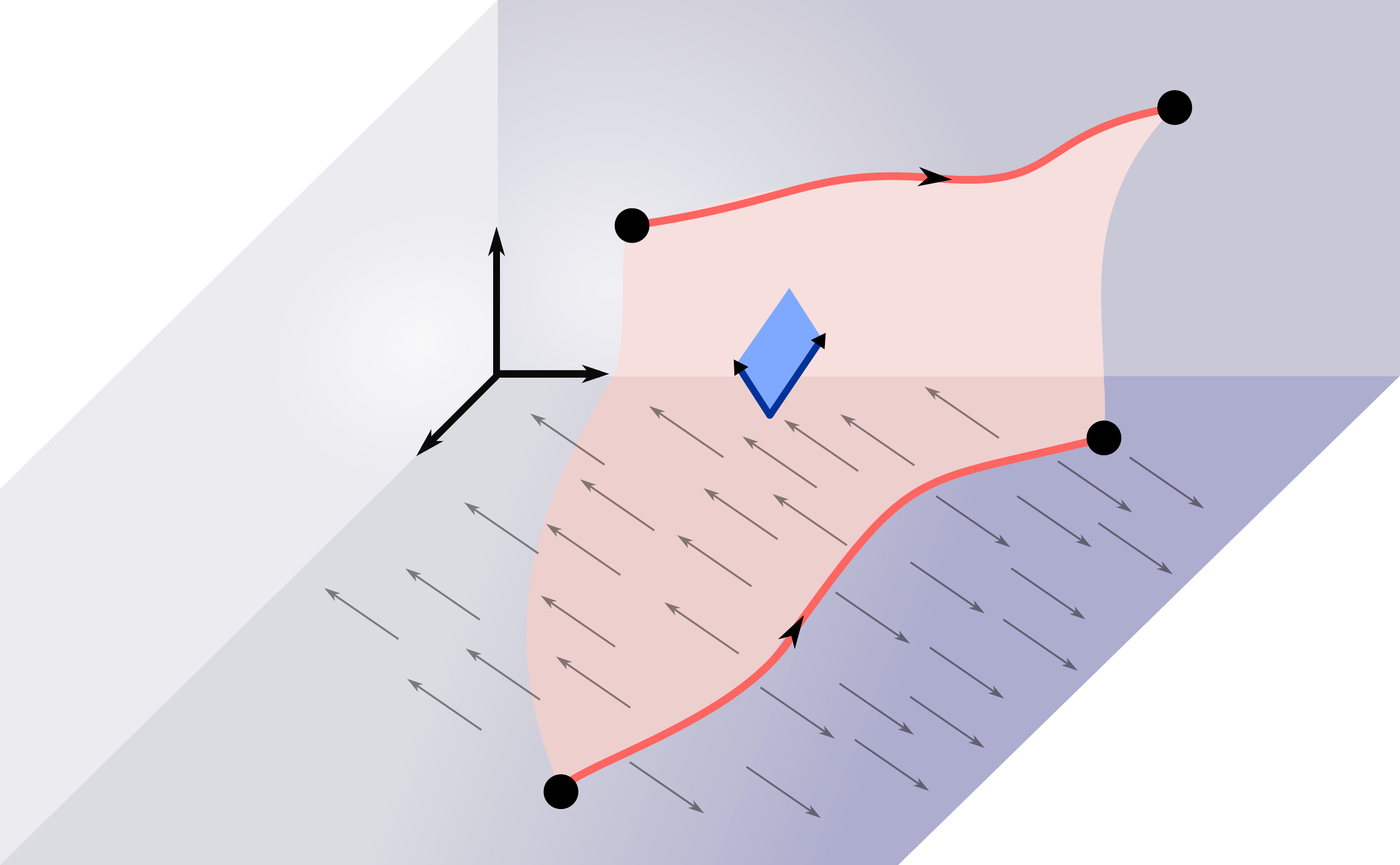}\label{fig:vortex worldsheet}}%
    \put(0.315,0.415){\color[rgb]{0,0,0}\makebox(0,0)[lb]{$t$}}%
    \put(0.40647646,0.37095953){\color[rgb]{0,0,0}\makebox(0,0)[lb]{$x$}}%
    \put(0.28,0.30){\color[rgb]{0,0,0}\makebox(0,0)[lb]{$y$}}%
    \put(0.60,0.38260568){\color[rgb]{0,0,0}\makebox(0,0)[lb]{$\kappa$}}%
    \put(0.49,0.34){\color[rgb]{0,0,0}\makebox(0,0)[lb]{$\lambda$}}%
  \end{picture}%
}
\caption{Vortex excitations in spacetime. \subref{fig:vortex worldline} A vortex particle in two spatial dimensions traces out a worldline in spacetime. It is parametrized by the line element $J^\mathrm{V}_\kappa(x)$. \subref{fig:vortex worldsheet} A vortex line in three spatial dimensions traces out a worldsheet in spacetime. Now we need two indices to define a surface element of the worldsheet, as such it is parametrized by the two-form field $J^\mathrm{V}_{\kappa\lambda}$. The vortices interact by exchanging gauge fields, that couple locally to the worldline ($b_\kappa$) or worldsheet ($b_{\kappa\lambda}$).}\label{fig:vortex worldline and worldsheet}
\end{figure*}
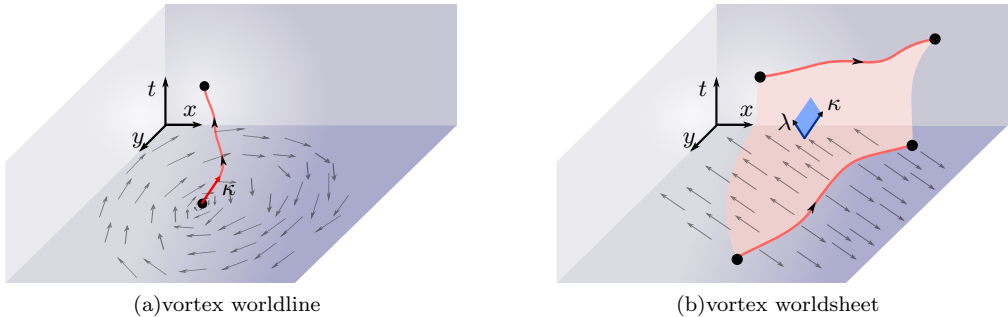
         
Until very recently it was not quite known how to formulate vortex duality in the natural 3+1 dimensions of the physical world. The obstruction was of a technical 
nature. In three space dimensions vortices are lines, which implies that in spacetime (as of relevance to the zero-temperature physics) vortices 
correspond to quantum strings (Fig. \ref{fig:vortex worldline and worldsheet}). Instead of the tangle of vortex worldlines in 2+1D,  the dual condensate now consists of a `foam' formed from the vortex worldsheets. Although unrelated to fundamental string theory, one cannot rely on the standard methods of quantum field theory for the description 
of such a condensate \cite{Rey89,Franz07} (see also Ref. \onlinecite{MotrunichSenthil05}). Its workings were tackled only recently \cite{BeekmanSadriZaanen11}. 
We will review this below, but the outcome is actually rather straightforward:
this `two-form Higgs phase' is qualitatively very similar to the standard (relativistic) superconductor, the main difference being in the counting of degrees of freedom.

Here we report on the extension from the neutral superfluid to the charged superconductor in 3+1D. Although far from self-dual, the charged Bose-Mott insulator as a `dual stringy superconductor' is behaving as a normal superconductor to the extent that the topology of the phase diagram of the 
Mott insulator in 3+1D is a `dual mirror image' of the phase diagram of a normal superconductor: our main result, Fig. \ref{fig:Mott phase diagram}. 
In a normal superconductor, the 
control parameters are temperature and applied magnetic field. As we will discuss in detail, the magnetic field dualizes into {\em applied current} in the Mott insulator,
and after re-identification of this axis the phase diagrams on both sides of the superconductor--Mott insulator transition acquire the same topology. To read off the physics of the Mott insulator, one can just depart from the standard wisdoms for superconductors using the `dual dictionary' summarized in Table \ref{table:SC MI correspondences}.

\begin{table}
  \begin{center}
\begin{ruledtabular}
\begin{tabular}{lclc}
 \multicolumn{2}{c}{superconductor} & \multicolumn{2}{c}{type-II Mott insulator} \\
 \cline{1-2} \cline{3-4}
 superfluid condensate & $| \Psi |^2$ & vortex condensate & $| \Phi|^2$ \\
 photon field & $\mathbf{A}$ & dual gauge field & $\mathbf{b}$\\
 applied magnetic field & $\mathbf{B}$ & applied current & $\mathbf{J}$ \\
 London penetration depth & $\lambda_\mathrm{L}$ & Mott proximity depth & $\lambda_\mathrm{M}$\\
 flux quantum & $\Phi_0$ & current quantum & $I_0$\\
 Meissner state & & insulating state& \\
 Abrikosov lattice & & current line lattice & \\
 electromagnetic vacuum & & superconductor & \\
 \emph{no dual}       &   & Maxwell vacuum & \\
\end{tabular}
\end{ruledtabular}
\end{center}
\caption{ Duality dictionary. The superconductor and the Mott insulator viewed as  vortex superconductor are quite similar, but
the meaning of the various physical quantities `turns upside down'. Magnetic fields turn into currents  with the ramification that 
the Abrikosov lattice of magnetic fluxes penetrating the superconductor turns into a lattice of current fluxes penetrating the Mott insulator, the type-II phase. The `mirror image' is not perfect (not self dual), which leaves more states
to play with in the case of the Mott insulator than for the normal superconductor (see Sec. \ref{sec:Phase diagram}).
 }\label{table:SC MI correspondences}
\end{table}

A quite counterintuitive prediction follows: upon reducing temperature, one will find a thermal phase transition to the Mott insulator with the same thermodynamical
($XY$) signatures as in a bosonic (local pair) superconductor. In a normal superconductor one applies magnetic fields to probe the  ``generalized rigidity'' \cite{Anderson84} of the ordered state. The dual of the magnetic field becomes in the Mott insulator the electric current. Just as in the Meissner phase the magnetic field is expelled, in the ``type-I Mott insulator'' the electric current is expelled (with an associated  dual penetration depth): this is just showing that the system is an insulator.

However, starting from local pairs with a very short coherence length, one is generically dealing with type-II behavior, both in the normal and dual superconductors. In duality language, the difference between type-I and type-II behavior is due to whether the disordering particles/strings (vortices) have net attractive or repulsive interactions. In a normal superconductor
a magnetic field that exceeds the lower critical field will penetrate in the form of an Abrikosov lattice of vortex lines, carrying each a quantized magnetic flux. The dual of the type-II superconductor is the ``type-II Bose-Mott insulator'' \footnote{The name ``type-II Mott insulator'' was used earlier by  Lee and Kivelson\cite{LeeKivelson03} to denote insulators that go continuously into a conductive regime
by doping. Here we condisder the type-II phenomenon of Abrikosov-like vortices in the
undoped, bosonic case.}: when the external current exceeds a ``lower critical current'' it will penetrate the Mott insulator in the form of  a lattice of `wires' carrying each a quantized supercurrent! Macroscopically it will behave just as a normal superconductor, which turns into a dissipative metallic state at the thermal transition where the dual order disappears. To find out whether such a superconductor is actually 
a Mott insulator in disguise one has to design experiments which are the `current analogues' of the decoration experiments that led to the discovery of
the  Abrikosov lattice. 

As an immediate application of the idea we suggest to search for type-II Mott insulation behavior in underdoped cuprate superconductors, widely believed to be dominated by phase fluctuations  \cite{EmeryKivelson95a}. Many researchers in the field are by now convinced that the so-called pseudogap regime consists of pre-formed Cooper pairs (bosons) that bind at a higher temperature $T^*$, whereas phase coherence and hence superconductivity set in only at at lower temperature $T_\mathrm{c}$. In this scenario, the transition from the superconducting to pseudogap phase is precisely of the $XY$-disordering type handled so well by vortex duality (see also Ref. \onlinecite{HerzogEtAl07}). Therefore we propose that in the vincinity of the quantum phase transition, the pseudogap phase is a type-II Bose-Mott insulator, and suggest several experimental setup that may verify the formation of quantized current lines in Fig. \ref{fig:Mott vortex experiments}.

The paper is organized as follows. We briefly recollect how the Bose-Hubbard model at zero chemical potential maps to the $XY$-model in Sec. \ref{sec:Bose-Hubbard model}. Also the well-established vortex dualization procedure for 2+1 dimensions is reviewed in Sec \ref{sec:Vortex duality in 2+1D}. Subsequently we will summarize the results of Ref. \onlinecite{BeekmanSadriZaanen11} in Sec. \ref{sec:Vortex duality in 3+1D}, where we derive the vortex duality in 3+1D as well, and show how to incorporate the electromagnetic field to model the superconductor in which we are interested. The remainder contains new results. From the vortex duality in 3+1D, the prediction of a dual Meissner effect for current and in particular quantized lines of electric current immediately follows in Sec. \ref{sec:Quantized current lines}. The main result is the phase diagram in Sec. \ref{sec:Phase diagram}. We propose several experiments that may confirm the existence of the quantized current lines in Sec. \ref{sec:Proposed experiments}
. The conclusions in Sec. \ref{sec:Conclusions} are followed by Appendix \ref{sec:Giant proximity effect}, which discusses the possible relevance of our findings to the so-called ``giant proximity effect'', and Appendix \ref{sec:Transformation properties of current worldsheets} regarding Lorentz transformations of vortex worldsheets.

\section{Bose-Hubbard model}\label{sec:Bose-Hubbard model}
Here we recall shortly the Bose-Hubbard model of bosons hopping on a hypercubic lattice in $D$ dimensions. For more detailed work see Refs. \onlinecite{FisherLee89,FisherEtAl89,BeekmanSadriZaanen11}. Let us first derive the relativistic continuum Ginzburg--Landau model of a superconductor by straightforward coarse graining. The Hamiltonian of the Bose-Hubbard model is,
\begin{equation}\label{eq:Bose-Hubbard Hamiltonian}
H_{\mathrm{BH}} = -\frac{t}{2}\sum_{\langle ij\rangle} (b^\dagger_i b^{\phantom{\dagger}}_j + b^\dagger_j b^{\phantom{\dagger}}_i) - \mu \sum_i n_i + \frac{U}{2} \sum_i (n_i-1)n_i.
\end{equation}
Here  $b^\dagger_i$ and $b^{\phantom{\dagger}}_i$ are boson creation and annihilation operators that satisfy the commutation relation $[b^{\phantom{\dagger}}_i , b^\dagger_j] = \delta_{ij}$. The number operator is $n_i = b^\dagger_i b^{\phantom{\dagger}}_i$. Furthermore, the energy scales are the boson hopping $t$, the on-site repulsion $U$ and the chemical potential $\mu$. We shall assume that the chemical potential is tuned so that there is an integer number $n_0$ of bosons per site (``zero chemical potential''). Then we can make a change of variables $b^\dagger_i  = \sqrt{n_0} \te^{\ti \varphi_i}$, so that the new conjugate variables satisfy the commutation relation $[\varphi_i,n_j ] = \ti \delta_{ij}$. Substituting this definition in Eq. (\ref{eq:Bose-Hubbard Hamiltonian}) leads to,
\begin{equation}
 H = -J \sum_{\langle ij \rangle}(1- \cos(\varphi_i - \varphi_j)) + \frac{U}{2} \sum_i (n_i - 1)n_i.
\end{equation}
Here we have defined $J = tn_0$ and added an unimportant constant. The physics of the weak and strong coupling limits is immediately clear: for large $J/U$, we have a superfluid where spatial fluctuations in the phase $\varphi$ are very costly. For small $J/U$ the on-site repulsion dominates and the bosons are confined to their lattice sites: the Mott insulator. 

For the  quantum field-theoretic formulation, we move from a Hamiltonian to a Lagrangian formalism, by noting that the canonical momentum is $\pi_j = \hbar n_j$, which leads to the Lagrangian by Legendre transformation (where $ 
\partial_t \varphi_j = \frac{\partial H}{\partial \pi_j}  = \frac{U}{\hbar^2} \pi_j$),
\begin{align}
 L &= \sum_i \pi_i \partial_t \varphi_i - H \nonumber\\
 &= \frac{\hbar^2}{2U} \sum_i (\partial_t \varphi_i)^2 - J  \sum_{\langle i,j\rangle} \big(1-\cos(\varphi_i - \varphi_j) \big).
\end{align}

Now we can take the continuum limit in $D$ space dimensions $ a^D\sum_i \mapsto \int \td^D x;  \ \   \varphi_i - \varphi_j \to a \nabla \varphi(x)$, where $a$ is the lattice constant. This leads to the partition function $Z = \te^{-\frac{1}{\hbar}S_\mathrm{E}}$ in imaginary time $t = \ti \tau$ where,
\begin{align}
  S_\mathrm{E} &= \frac{1}{a^D} \int \td \tau \td^D x\ \big[ -\frac{\hbar^2}{2U}(\partial_\tau \varphi)^2 - \frac{J}{2}a^2(\nabla \varphi)^2 \big] \nonumber\\
  &\equiv \int \td \tau \td^D x \ \frac{1}{2}J a^{2-D}\ \big[-\frac{1}{c^2_\mathrm{ph}} (\partial_\tau \varphi)^2 - (\nabla\varphi)^2 \big]. 
\end{align}
This is to be compared with the quantum action for a superfluid [cf. Eq. (3.13) in Ref. \onlinecite{FisherEtAl89}],
\begin{equation}\label{eq:superfluid action}
 S_\mathrm{E} = \int \td \tau \td^D x\ \big[ -\frac{1}{2} \hbar^2 \kappa (\partial_\tau \varphi)^2 - \frac{1}{2} \hbar^2 \frac{\rho_\mathrm{s}}{m} (\nabla \varphi)^2 \big].
\end{equation}
Hence we identify the compressibility $\kappa = \frac{1}{Ua^D}$, the superfluid density divided by the boson mass $\frac{\rho_\mathrm{s}}{m} = \frac{Ja^{2-D}}{\hbar^2}$ and the superfluid velocity $c_\mathrm{ph} = \frac{a}{\hbar} \sqrt{UJ}$. The energy scale $\sqrt{UJ}$ will play an important role in the discussion of the quantum of electric current later on. Defining the covariant derivative $\partial^\mathrm{ph}_\mu = (\frac{1}{c_\mathrm{ph}} \partial_\tau, \nabla)$, we find a convenient form of the action,
\begin{equation}
 S_\mathrm{E} = \int \td \tau \td^D x \ -\frac{1}{2}J a^{2-D}\  (\partial^\mathrm{ph}_\mu \varphi)^2.
\end{equation}

We are interested in charged superfluids, i.e. superconductors where the bosons must couple minimally to the electromagnetic potential, or photon field. Recall that the gauge-covariant derivative acts on the superfluid order parameter, which is a complex scalar field $\Psi = \sqrt{\rho_\mathrm{s}}\te^{\ti\varphi}$. Hence, the minimal coupling prescription in the London limit ($\rho_\mathrm{s}$ constant) is,
\begin{equation}
 |\partial^\mathrm{ph}_\mu \Psi|^2 \to |(\partial^\mathrm{ph}_\mu - \ti \frac{e^*}{\hbar} A^\mathrm{ph}_\mu)\Psi|^2 = \rho_\mathrm{s} (\partial^\mathrm{ph}_\mu \varphi - \frac{e^*}{\hbar} A^\mathrm{ph}_\mu)^2.
\end{equation}
Here $e^*$ is the electric charge of one boson (one Cooper pair). To preserve gauge invariance, the temporal component of the gauge potential should have the same velocity factor as the derivative, and therefore we define $A^\mathrm{ph}_\mu = (- \ti \frac{1}{c_\mathrm{ph}} V , \mathbf{A})$. In addition the Maxwell action for the dynamics of the electromagnetic field is included, which is governed by the speed of light $c$. Defining the electromagnetic field tensor $F_{\mu\nu} = ( \partial_\mu A_\nu - \partial_\nu A_\mu)$ where $\partial_\mu = ( \frac{1}{c} \partial_\tau ,\nabla)$ and $A_\mu = ( -\ti \frac{1}{c}V,\nabla)$, the total action is,
\begin{equation}
 S_\mathrm{E} = \int \td \tau \td^D x \ \big[-\frac{1}{2}J a^{2-D}\  (\partial^\mathrm{ph}_\mu \varphi - \frac{e^*}{\hbar} A^\mathrm{ph}_\mu)^2 - \frac{1}{4\mu_0} F_{\mu\nu}^2\big].\label{eq:dimensionful superconductor action}
\end{equation}
This is the relativistic Ginzburg--Landau (Abelian-Higgs) model, where we have suppressed the potential terms $\sim \alpha |\Psi|^2 + \beta | \Psi |^4$, which are frozen out in the London limit of small amplitude fluctuations, as we will assume throughout.

\section{Vortex duality in 2+1D}\label{sec:Vortex duality in 2+1D}
In the previous section we derived the weak-coupling continuum limit of the Bose-Hubbard model in terms of the  dynamics of the phase $\varphi$. It can also capture the strong-coupling phase if we incorporate the agents that destroy phase coherence: the vortices, windings of $2\pi$ of the phase field. From the dual viewpoint, vortices are particles, that can condense just as well as bosons can\cite{FisherLee89,Kleinert08,HerbutTesanovic96,CvetkovicZaanen06a,NguyenSudbo99,HoveSudbo00}. The vortex condensate corresponds to the state where the original variables $\varphi$ have completely lost their meaning. In other words, the weak-coupling phase of the vortices is the strong-coupling phase of the original variables.

It is useful to go over to dimensionless variables denoted by a prime,
\begin{equation}
 S_\mathrm{E} = \hbar S'_\mathrm{E},\quad x = ax', \quad\tau= \frac{a}{c_\mathrm{ph}}\tau',\quad A_m = \frac{\hbar}{e^*a} A'_m.
\end{equation}
We shall suppress the primes in the remainder. The dimensionless version of the action Eq. (\ref{eq:dimensionful superconductor action}) is $S_\mathrm{E} = \int \td \tau \td^D x \  {\cal L}$ with,
\begin{equation}\label{eq:dimensionless superconductor Lagrangian}
 {\cal L}= - \frac{1}{2g} (\partial^\mathrm{ph}_\mu \varphi - A_\mu)^2 - \frac{1}{4\mu} F_{\mu\nu}^2.
\end{equation}
Here the dimensionless coupling constants are,
\begin{equation}\label{eq:dimensionless coupling constants}
 \frac{1}{g}  = \frac{Ja}{\hbar c_\mathrm{ph}} , \qquad \frac{1}{\mu}  = \frac{\hbar a^{D-3}}{\mu_0 c_\mathrm{ph} {e^*}^2}.
\end{equation}

Two quantities are of interest in the duality. The first is the current $w_\mu =  \frac{1}{g} (\partial^\mathrm{ph}_\mu \varphi - A^\mathrm{ph}_\mu)$, related to the charged supercurrent as $w_\mu = \frac{\hbar}{e^*} J^\mathrm{EM}_\mu$. Then Eq. (\ref{eq:dimensionless superconductor Lagrangian}) can be dualized  by direct substitution into,
\begin{equation}\label{eq:dual superconductor Lagrangian}
{\cal L}_\mathrm{dual} =  \frac{1}{2}g w_\mu^2 - w_\mu (\partial^\mathrm{ph}_\mu\varphi - A^\mathrm{ph}_\mu) - \frac{1}{4\mu} F^2_{\mu\nu}.
\end{equation}

The second quantity of interest describes the Abrikosov vortices, which are singularities in the phase field $\varphi$. For the remainder of this section we specialize to 2+1 dimensions, for simplicity. Splitting the phase field into a smooth and a singular part, $\varphi = \varphi_\mathrm{smooth} + \varphi_\mathrm{sing}$, a vortex solution of winding number $N$ satisfies,
\begin{align}\label{eq:winding number}
 2\pi N &= \oint_{\partial {\cal S}} \td x_\mu \ \partial_\mu \varphi = \int_{\cal S}\td S_\kappa \ \epsilon_{\kappa\nu\mu} \partial_\nu \partial_\mu ( \varphi_\mathrm{smooth} + \varphi_\mathrm{sing}) \nonumber \\
 &=  \int_{\cal S}\td S_\kappa \ \epsilon_{\kappa\nu\mu} \partial_\nu \partial_\mu \varphi_\mathrm{sing},
\end{align}
by Stokes' theorem. The derivatives acting on a singular field do not commute.

\subsection{The superconductor is a Coulomb gas of vortices}
On the smooth part we can perform integration by parts in Eq. \eqref{eq:dual superconductor Lagrangian}, to obtain a term $(\partial^\mathrm{ph}_\mu w_\mu) \varphi_\mathrm{smooth}$, and $\varphi_\mathrm{smooth}$ can be integrated out as a Lagrange multiplier for the contraint $\partial^\mathrm{ph}_\mu w_\mu =0$, the conservation of supercurrent (continuity equation) in the superconductor. In 2+1 dimensions, this constraint can be explicitly enforced by expressing it as the curl of a dual gauge field,
\begin{equation}
 w_\mu = \epsilon_{\mu\nu\kappa} \partial^\mathrm{ph}_\nu b_\kappa.
\end{equation}
This expression is invariant under the addition of the gradient of any smooth scalar field,
\begin{equation}
  b_\kappa (x) \to b_\kappa(x) + \partial_\kappa \varepsilon(x).
\end{equation}

Substituting this definition in Eq. (\ref{eq:dual superconductor Lagrangian}) leads to,
\begin{align}\label{eq:dual gauge field Lagrangian}
{\cal{L}}_\mathrm{2+1d} &=
  \frac{1}{2}g(\epsilon_{\mu\nu\kappa} \partial^\mathrm{ph}_\nu b_\kappa)^2
  -b_\kappa J^\mathrm{V}_\kappa \nonumber \\
&\phantom{=} + \epsilon_{\mu\nu\kappa} \partial^\mathrm{ph}_\nu b_\kappa A^\mathrm{ph}_\mu 
  - \frac{1}{4\mu} F^2_{\mu\nu}.
\end{align}
Here we have performed integration by parts on the second term, and we recognize the expression from Eq. (\ref{eq:winding number}). Therefore we define $J^\mathrm{V}_\kappa (x) = \epsilon_{\kappa \nu \mu} \partial_\nu \partial_\mu \varphi_\mathrm{sing} (x)$ as the \emph{vortex current}. It is a one-form (vector) field because a vortex point particle traces out a worldline, with line element $J^\mathrm{V}_\kappa (x)$. Then from the coupling term $b_\kappa J^\mathrm{V}_\kappa$ in Eq. (\ref{eq:dual gauge field Lagrangian}) we see that vortices interact by exchanging dual gauge particles $b_\kappa$. In other words: the 2+1D neutral superfluid ($e^* \to 0$) is a Coulomb vacuum for the vortices with long-range interactions mediated via dual `photons' $b_\kappa$.

In the charged superfluid, the current $w_\mu = \epsilon_{\mu\nu\kappa}  \partial^\mathrm{ph}_\nu b_\kappa$ also couples to the real electromagnetic photon $A_\mu$, rendering the interaction between Abrikosov vortices short-ranged. However, for the formation of the vortex condensate as described below, it is unimportant whether or not it is coupled to electromagnetic field. This implies that the extension to the charged superfluid/superconductor is straightforward by choosing $e^* > 0$.

\subsection{The Bose-Mott insulator is a dual superconductor}
The true power of the duality lies in the fact that the strong-coupling phase, i.e. wildly fluctuating phase fields, can be described as an ordered state in terms of the vortices. Vortex--antivortex pairs can spontaneously emerge and annihilate in the form of closed spacetime loops. In the Coulomb phase (superfluid), such processes are heavily suppressed, as the coupling constant $g$ acts as the line tension of such spacetime loops.

\begin{figure*}
\begin{center}
\includegraphics[width=15cm]{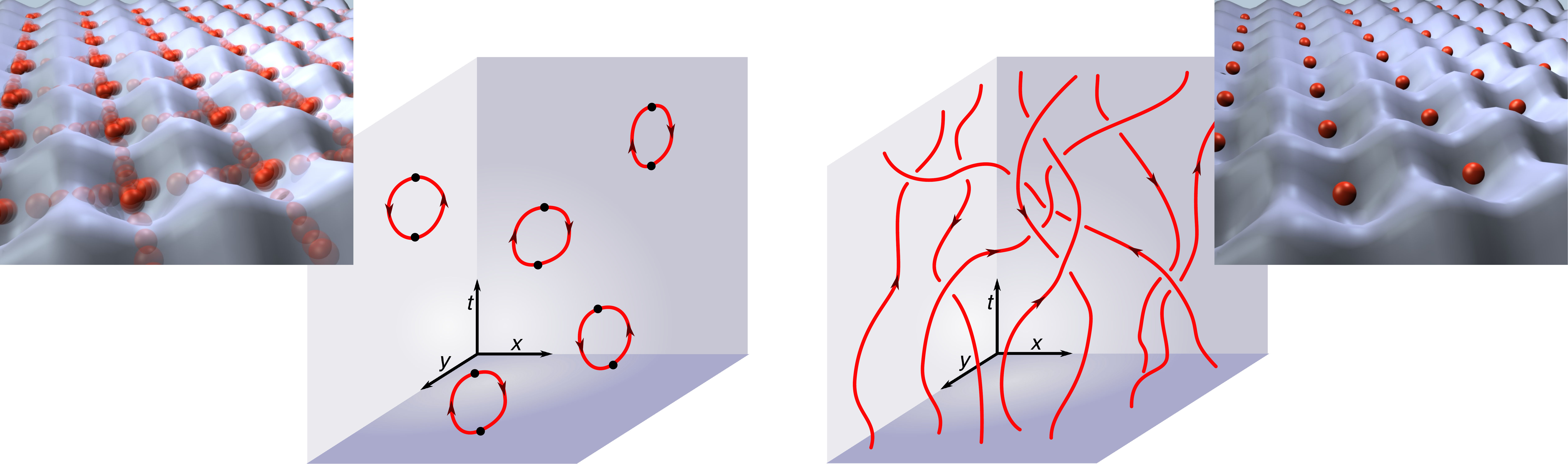}
\caption{Cartoon of the superconductor--Bose-Mott insulator transition in the dual vortex-language,  for simplicity shown in 2+1 dimensions. In the superconducting state of a system of bosons tunneling between potential wells, all the particles are completely delocalized (left). When the local repulsions
become large enough at a density of one boson per well a ``trafffic jam'' sets in and the Bose-Mott insulator is formed (right).  In the dual view one focuses on the physics of the 
vortices, the topological excitations of the superconducting state. In the superconductor these occur as virtual fuctuations of bound vortex--antivortex pairs, forming closed
loops of worldlines in spacetime (red lines, middle left). However, at the quantum phase transition to the Mott insulator these loops ``blow out''; the Mott insulator
itself corresponds to a tangle of worldlines (middle right). Due to long-range interactions between vortices this just represents a relativistic superconductor (Higgs vacuum) of vortices,
the Higgs mass being coincident with the Mott gap. Here we consider the situation in 3+1D where the vortices turn into strings. The Mott insulating state is now dually described as a `foam of vortex strings' in spacetime, eventually behaving just as a three dimensional superconductor.}\label{fig:vortex duality}
\end{center}
\end{figure*}

As the coupling constant $g$ decreases, the vortex worldlines grow in size and number, until at the critical point $g_\mathrm{c}$ they span the whole system. At that point, vortices and antivortices can be created energetically for free. In other words, across the phase transition we find an ordered state, the vortex condensate $\Phi$, out of which vortices can be pulled everywhere for free, just as Cooper pairs can be pulled out of the superconducting condensate. This is pictured in Fig. \ref{fig:vortex duality}.

Since the dual gauge fields $b_\kappa$ couple to the vortex condensate $\Phi$ just as the electromagnetic field $A_\mu$ couples to the superconducting condensate $\Psi$, the vortex condensate is a dual superconductor. We know what the phase transition implies for massless gauge fields: as they couple minimally to a condensate field $\Phi = | \Phi | \te^{\ti \phi}$ they become massive due to the Anderson--Higgs mechanism \cite{Kleinert08, FisherLee89, FradkinShenker79,HerbutTesanovic96, CvetkovicZaanen06a, NguyenSudbo99,HoveSudbo00,Zee00,Fisher04}. We end up with the dual superconductor in terms of the gauge field $b_\kappa$,
\begin{align}\label{eq:2+1D dual GL Lagrangian}
{\cal{L}}_\mathrm{2+1d}& =
  \frac{1}{2}g(\epsilon_{\mu\nu\kappa} \partial^\mathrm{ph}_\nu b_\kappa)^2 + \epsilon_{\mu\nu\kappa} \partial^\mathrm{ph}_\nu b_\kappa A^\mathrm{ph}_\mu - \frac{1}{4\mu} F^2_{\mu\nu}  \nonumber \\
  & \phantom{=} + \frac{1}{2} | (\partial^\mathrm{ph}_\kappa - \ti b_\kappa)\Phi|^2 - \frac{\tilde{a}}{2} | \Phi|^2 - \frac{\tilde{\beta}}{4}| \Phi|^4,
\end{align}
where $\tilde{\alpha}$ and $\tilde{\beta}$ are dual Ginzburg--Landau parameters. The gauge field $b_\kappa$ obtains a Higgs mass $|\Phi|^2/g$, and furthermore the longitudinal polarized photon now becomes a physical degree of freedom. This is the dual way of expressing that the single massless zero sound mode of the superfluid turns into the two gapped modes (doublon and holon) in the Mott insulator\cite{CvetkovicZaanen06a}! Therefore the former is referred to as the Coulomb phase and the latter as the Higgs phase in terms of the vortex operators. 

In the charged superconductor, the original Anderson--Higgs-massive Goldstone mode (phase of the superconducting condensate $\Psi$) dualizes into $b_\kappa$ and its gets contributions to its mass both from the electromagnetic field and from the vortex condensate. Furthermore the vortex condensate $\Phi$ gives rise to an additional mode. In other words, the transverse polarization $b_\mathrm{T}$ is the superconductor sound mode, and the longitudinal polarization $b_\mathrm{L}$ is the vortex condensate sound mode. They are gapped and 
degenerate. In addition there is still the electromagnetic field $A_\mu$ which is also gapped. To avoid confusion by all the various gauge fields in this `dual equation', it is often useful to first regard the neutral limit $e^* \to 0$ and subsequently let the electromagnetic field enter weakly coupled to the current $w_\mu$.

\section{Vortex duality in 3+1D}\label{sec:Vortex duality in 3+1D}
The question is now how this generalizes to higher dimensions. In the boson language, the superfluid/Mott insulator picture is unaltered. But vortices become extended objects: Nielsen--Olesen (non-critical) strings in 3+1 dimensions\cite{NielsenOlesen73}. The Abrikosov vortex line traces out a worldsheet in spacetime, with surface element $J^\mathrm{V}_{\kappa\lambda} = \epsilon_{\kappa\lambda \nu \mu} \partial_\nu \partial_\mu \varphi_\mathrm{sing}$ (Fig. \ref{fig:vortex worldline and worldsheet}). The temporal components $J^\mathrm{V}_{tl}$ are the density of the vortex line along $l$, while $J^\mathrm{V}_{kl}$ denotes the motion in the direction $k$ of the line along $l$, with continuity equations $\partial_\kappa J^\mathrm{V}_{\kappa\lambda}= 0$ for all $\lambda$.

\subsection{The Coulomb phase}
The $J^\mathrm{V}_{\kappa\lambda}$ are two-form antisymmetric tensor currents\cite{Naber97}, and the dual gauge fields mediating the interaction between vortices also become two-form fields $b_{\kappa\lambda}$. For the Coulomb phase (superfluid), the generalization is straightforward. The conservation of supercurrent $\partial^\mathrm{ph}_\mu w_\mu =0$ can be enforced by expressing it as the curl of this two-form gauge field,
\begin{equation}\label{eq:two-form gauge field definition}
  w_\mu = \epsilon_{\mu\nu\kappa\lambda} \partial^\mathrm{ph}_\nu b_{\kappa\lambda}.
\end{equation}
Here $\epsilon_{\mu\nu\kappa\lambda}$ is the completely antisymmetric Levi-Civita tensor in four dimensions. This expression is invariant under the addition of the gradient of any smooth vector field,
\begin{equation}
 b_{\kappa\lambda} \to b_{\kappa\lambda} + \partial_\kappa \varepsilon_\lambda -\partial_\lambda \varepsilon_\kappa.
\end{equation}
Even though the transformation looks like the electromagnetic field strength $F_{\mu\nu} = \partial_\mu A_\nu - \partial_\nu A_\mu$, it should not be confused with the actual field strength related to this dual gauge field, which is $w_\mu$ in Eq. \eqref{eq:two-form gauge field definition}. The gauge transformation itself has a redundancy, as another gauge transformation $\varepsilon'_\lambda = \varepsilon_\lambda + \partial_\lambda \eta$ would yield the exact same transformation, which is sometimes referred to as ``gauge in the gauge'', and is important for the counting of degrees of freedom \cite{HenneauxTeitelboim92}.

The Lagrangian of the Coulomb phase Eq. \eqref{eq:dual gauge field Lagrangian} generalizes to
\begin{align}
 {\cal{L}}_\mathrm{3+1d} &=
  \frac{1}{2}g(\epsilon_{\mu\nu\kappa\lambda} \partial^\mathrm{ph}_\nu b_{\kappa\lambda})^2
  -b_{\kappa\lambda}J^\mathrm{V}_{\kappa\lambda} \nonumber \\
&\phantom{=} + \epsilon_{\mu\nu\kappa\lambda} \partial^\mathrm{ph}_\nu b_{\kappa\lambda} A^\mathrm{ph}_\mu 
  - \frac{1}{4\mu} F^2_{\mu\nu}.
\end{align}

Again, the vortices have long-range interactions mediated by the superfluid sound mode, now represented by $b_{\kappa\lambda}$. In the neutral case $e^* \to 0$, we retrieve the theory of a free and massless two-form gauge field in 3+1D, which is known to have one propagating degree of freedom\cite{HenneauxTeitelboim92}. It is the purely transversal component with $\kappa$ and $\lambda$ each taking a transversal direction. For the electrically charged case this mode becomes gapped, just as those of the electromagnetic field $A_\mu$ do. Nevertheless, the superfluid dualizes into a gas of vortex worldsheets interacting via two-form gauge fields. Vortex--antivortex creation and annihilation events (quantum fluctuations in the superfluid) take the form of small closed worldsheet surfaces, suppressed by the large coupling constant $g$.

\subsection{The Higgs phase}
The dual condensate corresponds with a `foam' formed by vortex strings filling 3+1D spacetime. This `stringy Higgs phase' is obviously somehow different from the conventional `particle' Higgs phase of 2+1D. Surely the superfluid is ordered in terms of phase dynamics, and the Mott insulator corresponds to the completely phase-disordered state. A phase winding of $2\pi$ corresponds to the local formation of a vortex excitation. Therefore we expect that the Mott insulator is again a condensate of such vortex excitations. This cannot be dealt with using standard field-theoretic techniques as in 2+1D, where one can write down a quantum field theory of meandering vortex worldlines, the collective field of which takes the form of a Ginzburg--Landau scalar field\cite{Kleinert08}. Now we should have a quantum field theory of vortex worldsheets: a string field theory. Such a theory is not yet available in closed form.

However, at least for the mundane finite-energy vortices in condensed matter---as opposed to the coreless critical strings of string theory---the final result must be the Bose-Mott insulator. This insulator has two gapped doublon and holon modes regardless the dimensionality of the system. Hence, whatever the vortex string condensate may be, it should add precisely one dynamic mode aside from the sound of the superfluid, and both modes should become gapped and degenerate. This is precisely the guiding principle we employed in our earlier work Ref. \onlinecite{BeekmanSadriZaanen11}, which we now briefly summarize. It turns out that earlier attempts at establishing a field theory of vortex strings\cite{MarshallRamond75,Rey89,Franz07} assumed that one can straightforwardly generalize the minimal coupling construction for the phase of the Higgs field of second quantization $(\partial_\mu \varphi - b_\mu)$ to the stringy case $(\partial_\mu c_\nu - \partial_\nu c_\mu - b_{\mu\nu})$. However, this implies that one 
has to associate a vectorial phase field $c_\mu$ to the string condensate, which yields two longitudinal photons and three massive modes in total. Although this might be accurate for critical strings, it does not add up to the doublet of gapped modes of the Bose-Mott insulator.

Therefore we reconsidered the status of the two-form gauge field $b_{\kappa\lambda}$. The single reason for introducing it in Eq. \eqref{eq:two-form gauge field definition} was that the supercurrent $w_\mu$ is a conserved quantity. If we resubstitute this definition in the Higgs Lagrangian for 2+1D Eq. \eqref{eq:2+1D dual GL Lagrangian}, we obtain,
\begin{align}\label{eq:2+1D supercurrent Higgs Lagrangian}
 {\cal{L}} &=  
    \frac{1}{2}g {w_\mu}^2 + \frac{1}{2}| \Phi |^2 w_\mu \frac{1}{\partial^2} w_\mu + w_\mu A_\mu  - \frac{1}{4\mu} F^2_{\mu\nu},
\end{align}
The second term explains the Anderson--Higgs mechanism in the sense that supercurrent can no longer be created for free (the modes are gapped/massive), but it does not explicitly demonstrate where the additional degree of freedom, the `longitudinal photon', originates.

We need to realize that a vortex is a source or sink of supercurrent. Therefore, in the vortex condensate where vortices can be created for free at every point in space, the conservation of supercurrent is violated. A more precise statement is that there is a superposition of having 0, 1, or any number of vortices at any point, such that correlations of the phase field vanish completely, a notion we explored further in Ref. \onlinecite{ZaanenBeekman12}. Hence, the constraint $\partial^\mathrm{ph}_\mu w_\mu = 0$ is removed, which liberates the longitudinal component of the current as a physical degree of freedom.

Therefore the Lagrangian Eq. \eqref{eq:2+1D supercurrent Higgs Lagrangian} is valid in any dimension. The vortex condensate amounts to the appearence of the second term $\sim |\Phi|^2$: the Higgs mass / condensate density / Mott gap. Concurrently, the supercurrent is no longer conserved, and the longitudinal component of the supercurrent enters as a physical degree of freedom, leading to two gapped modes in the Bose-Mott insulator, in any dimension. 

The electromagnetic field couples as always to the electric current $J^\mathrm{EM}_\mu = e^* w_\mu$. 

Summarizing, the 3+1D Bose-Mott insulator is again a dual superconductor, albeit of a special kind where two-form gauge fields take the role of Higgsed photons. Nevertheless, the dual order parameter $\Phi$ instigates a dual Meissner effect by causing electric current to decay exponentially resulting in the insulating behavior. It also immediately suggests that the vortex condensate has vortices $\mathcal{J}^\mathrm{V}_{\kappa\lambda} = \epsilon_{\kappa\lambda\mu\nu}\partial^\mathrm{ph}_\mu \partial^\mathrm{ph}_\nu \phi$ of its own, which are lines of quantized electric current just as Abrikosov vortices are lines of quantized magnetic flux. This we will investigate in further detail below.

\subsection{Dual vortices and the dual gauge field}
Still, the question remains how Eq. \eqref{eq:2+1D supercurrent Higgs Lagrangian} can be expressed in terms of the two-form gauge field $b_{\kappa\lambda}$. This issue is particularly important considering vortices in the dual condensate. What are the singularities in the phase field $\phi$ of the stringy vortex condensate order parameter $\Phi = | \Phi | \te^{\ti \phi}$?

All along the problem is how to match the gradient of this phase field to the two-form gauge field. What is the form of the minimal coupling analogous to $(\partial_\mu - \ti b_\mu)\Phi$  of  Eq. \eqref{eq:2+1D dual GL Lagrangian}:
\begin{equation}\label{eq:minimal coupling question}
 (\partial_\mu -\ti\  ???\ b_{\kappa\lambda}) \Phi \ ?
\end{equation}

Earlier work\cite{MarshallRamond75,Rey89,Franz07} proposed a vectorial phase field $\phi \to c_\nu$, but this implied a too large number of degrees of freedom as we already mentioned above. In our preceding work\cite{BeekmanSadriZaanen11} we proposed,
\begin{equation}
 (\partial_\mu - \ti \epsilon_{\mu\parallel\kappa\lambda} b_{\kappa\lambda})\Phi,
\end{equation}
where the three free indices in $\epsilon_{\mu\parallel\kappa\lambda}$ take values orthogonal to the four-momentum $\ti \partial_\mu \to p_\mu$ only. Essentially this corresponds to the generalized Lorenz gauge fix $\partial_\kappa b_{\kappa\lambda} = 0 \ \forall \lambda$. This does lead to the correct form Eq. \eqref{eq:2+1D supercurrent Higgs Lagrangian}. However, the relation between the phase field $\phi$ and the dual gauge field $b_{\kappa\lambda}$ is obscured, because cross-terms are explicitly eliminated by this gauge fix. The same thing happens in the Ginzburg--Landau equations in the Lorenz gauge fix, the phase and the photon fields become decoupled. Any inquiry into the vortex excitations, which are singularities in the phase field, cannot rely upon such gauge-fixed expressions.

It turns out that the issue of worldsheets of current lines is actually a surprisingly intricate affair. One would think that for instance a simple current-carrying copper wire cannot hold any macroscopic secrets, but the equations of motion when such a wire is regarded as a worldsheet in spacetime have never been established. In an earlier work\cite{BeekmanZaanen11} we performed such an investigation for the case of Abrikosov vortices in superconductors. Here everything fits together neatly. Start out with the relativistic equation of motion derived from Eq. 
\eqref{eq:dimensionful superconductor action}, 
\begin{equation}
 - \partial_\mu F_{\mu\nu} + A_\nu = \partial_\nu \varphi,
\end{equation}
where we suppressed all dimensionful constants. Now act on this expression with $\epsilon_{\kappa\lambda\rho\nu}\partial_\rho$ to find,
\begin{equation}\label{eq:Abrikosov vortex EoM}
 \partial^2 \epsilon_{\kappa\lambda \mu\nu} F_{\mu\nu} +\epsilon_{\kappa\lambda \mu\nu} F_{\mu\nu} = J^\mathrm{V}_{\kappa\lambda}.
\end{equation}
Here $J^\mathrm{V}_{tl}$ is the density of an Abrikosov vortex line in direction $l$, and  $J^\mathrm{V}_{kl}$ is the motion in direction $k$ of a such a vortex along $l$. By neglecting the first term which denotes the decay of the electromagnetic fields away from the vortex due to the Meissner effect, we see that a vortex line $J^\mathrm{V}_{tl}$ induces or couples to magnetic field $B_l = \epsilon_{kmn}F_{mn}$, and motion of the vortex induces electric field $J^\mathrm{V}_{kl} \sim \epsilon_{kltn} F_{tn} = \epsilon_{kln} E_n$, corresponding to the relation $\mathbf{v} \times \mathbf{B} = \mathbf{E}$, well known in vortex dynamics. 

\begin{figure*}
 \subfigure[static vortex line]{\includegraphics[width=3.0cm]{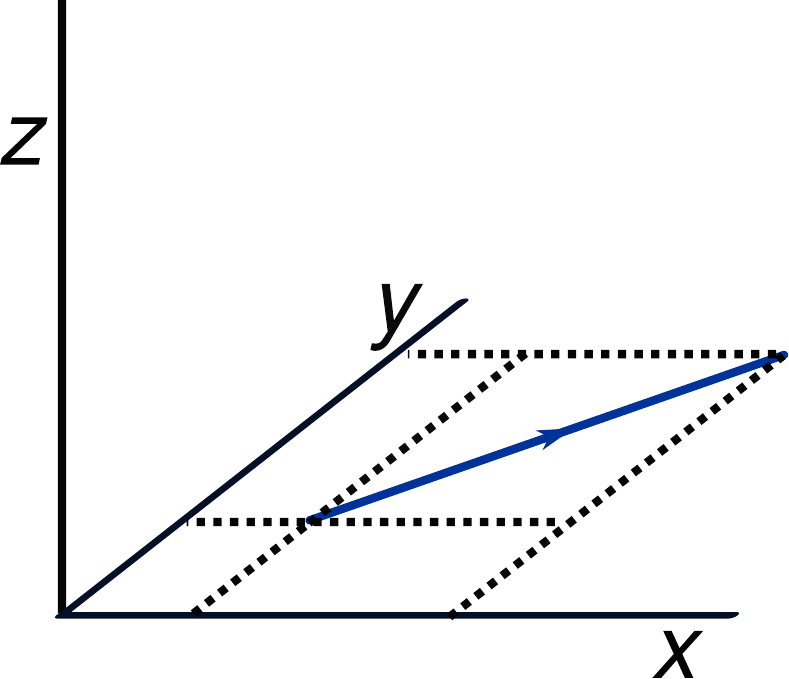}\label{fig:static vortex line}}
\hfill
\subfigure[dynamic vortex pancake]{\includegraphics[width=3.0cm]{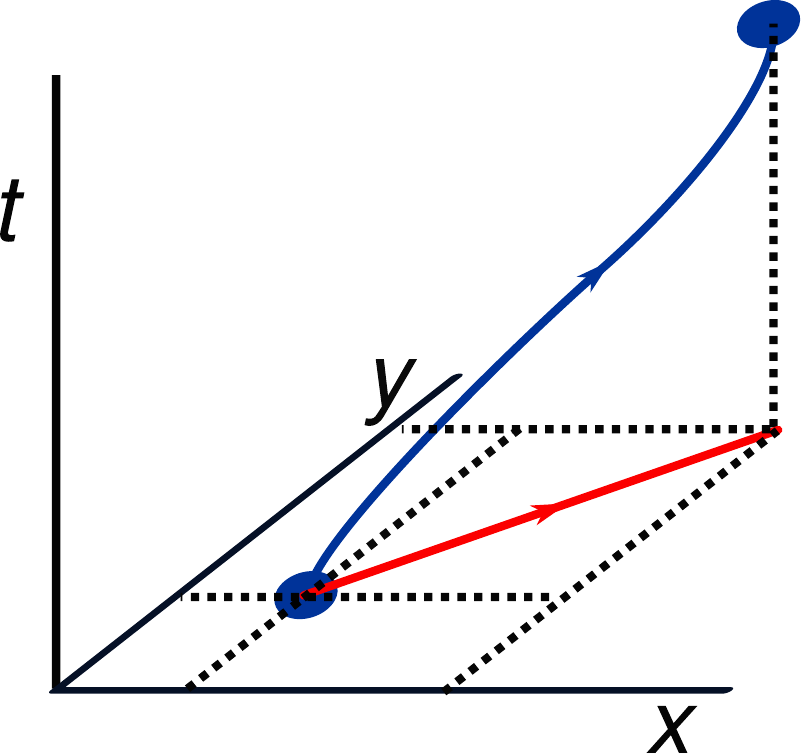}\label{fig:dynamic vortex pancake}}
\hfill
 \subfigure[static vortex worldsheet]{\includegraphics[width=3.0cm]{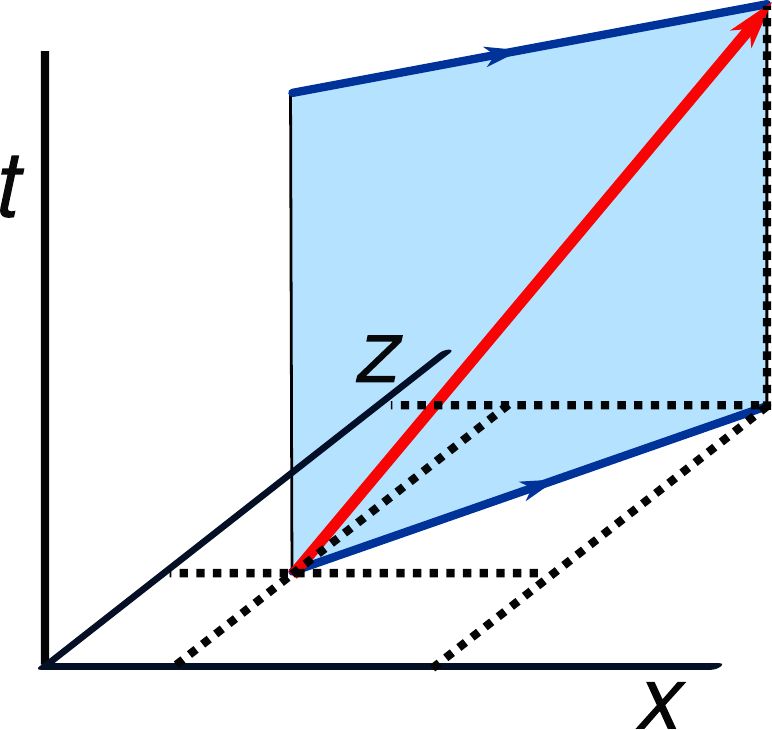}\label{fig:static vortex sheet}}
\hfill
 \subfigure[dynamic vortex worlsheet]{\includegraphics[width=3.0cm]{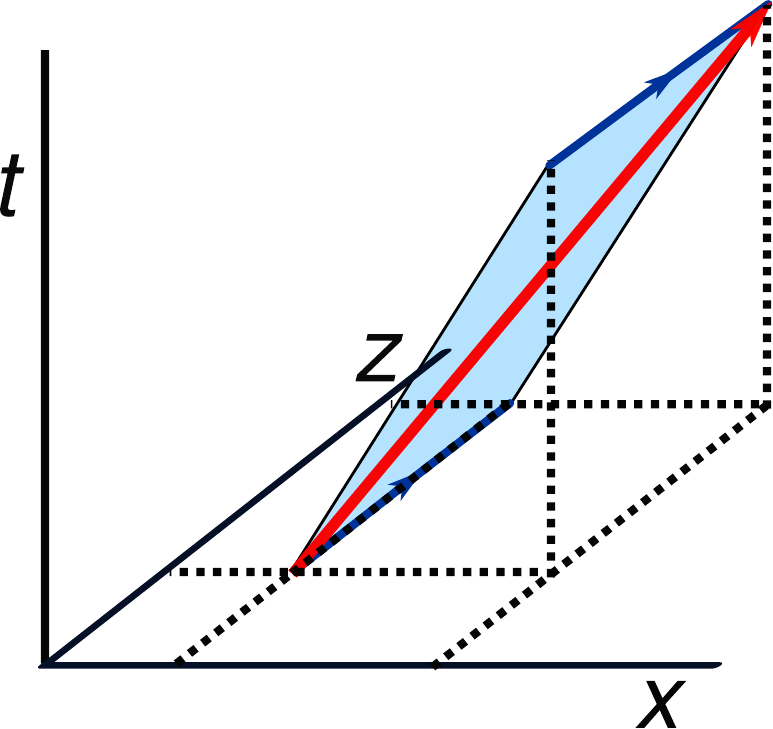}\label{fig:dynamic vortex sheet}}
\caption{\subref{fig:static vortex line} Static vortex line in the $xy$-plane; the current flows through the line. \subref{fig:dynamic vortex pancake} Vortex pancake moving in time (blue). The associated current in the spatial direction is shown in red. \subref{fig:static vortex sheet} Static vortex line in the $xz$-plane moving straight up in time. \subref{fig:dynamic vortex sheet} A vortex line in the $z$-direction moving in the $x$-direction through time. The last two worldsheet configurations correspond to the same electromagnetic current (red).
\label{fig:current world sheets}}
\end{figure*}

The reason why this works so well is that the two-form vortex source $J^\mathrm{V}_{\kappa\lambda}$ couples directly to the two-form electromagnetic tensor $F_{\mu\nu}$. Conversely, the electric current $w_\mu$ is a vector quantity. So we face a problem similar to Eq. \eqref{eq:minimal coupling question}, matching quantities of differrent geometrical nature. A closed-form relativistic expression is not found, but the physics at play can nevertheless be accurately comprehended. This is illustrated in Fig. \ref{fig:current world sheets}: imagine a volume element of current at position $\mathbf{x}$. This current element cannot distinguish between moving along a static line (wire) in a certain direction $l$, or being dragged along a vortex line along $k$ but  moving in direction $l$. Therefore, the current in direction $l$ gets contributions both from static lines pointing along $l$, denoted by $\mathcal{J}^\mathrm{V}_{tl}$, and from lines pointing in different directions $k$ moving in direction $l$, denoted by $\mathcal{J}^\mathrm{V}_{lk}$. Our task is to, given a vortex worldsheet $J^\mathrm{V}_{\kappa\lambda}$, derive the resulting electric current $w_\mu$.

It is instructive to first observe the 3+0D limit of static vortex lines carrying stationary current. This is in fact also the starting point for discussions of Abrikosov vortices. In the static limit the only vortex components are $\mathcal{J}^\mathrm{V}_l \equiv \mathcal{J}^\mathrm{V}_{tl} = \epsilon_{tlmn} \partial^\mathrm{ph}_m \partial^\mathrm{ph}_n \phi$, and the only dual gauge field components are $b_l \equiv b_{tl}$. The Lagrangian is Eq. \eqref{eq:2+1D dual GL Lagrangian} where however all indices denote spatial dimensions. It contains the standard form of the minimal coupling, and hence the equation of motion is readily obtained by variation with respect to $b_l$,
\begin{equation}
 g \epsilon_{lnm}\partial_n w_m + B_l + |\Phi|^2 b_l = |\Phi|^2 \partial_l \phi.
\end{equation}
Here $B_l = \epsilon_{lnm}\partial_n A_m$ is the magnetic field. Acting on this expression with $\epsilon_{rsl} \partial_s$ gives,
\begin{equation}\label{eq:dimensionless 3+0D EoM}
 g(\partial_r\partial_m - \delta_{rm}\partial^2)w_m +|\Phi|^2w_r + \epsilon_{rsl} \partial_s B_l =  |\Phi|^2 \mathcal{J}^\mathrm{V}_r.
\end{equation}
Here we clearly see that a static vortex line $ \mathcal{J}^\mathrm{V}_r$ sources electric current $w_r$ in the parallel direction. This current decays exponentially in the insulator. The current also sources the curl of magnetic field, as in the Amp\`ere--Maxwell equation but we will see later that this effect is very weak.

This is the situation of Fig. \ref{fig:static vortex line}. It is clear that by incorporating the time dimension, the vortex line traces out a worldsheet, the four-current has the charge density as temporal component, and hence Fig. \ref{fig:static vortex sheet} describes precisely the same situtation as Fig. \ref{fig:static vortex line}. The worldsheet surface element is spanned by the temporal component $w_t = c\rho$ and the spatial current $w_l$, such that the four-current is `the diagonal' of the worldsheet. Now consider the situation of Fig. \ref{fig:dynamic vortex sheet}. Here the worldsheet is `tilted' and therefore contains, next to the density $\mathcal{J}^\mathrm{V}_{tz}$, a component $\mathcal{J}^\mathrm{V}_{xz}$. This surface element is spanned by the two current components $w_x$ and $w_x$.

These notions are most conveniently expressed by using differential forms (see any text book on differential geometry, e.g. Ref. \onlinecite{Naber97}). Here $\td x^\mu$ represents a line element in direction $x^\mu$. A vector field $\mathsf{a}(x)$ is expanded in components as $\mathsf{a}(x) = \sum_\mu a_\mu(x) \td x^\mu \equiv a_\mu(x) \td x^\mu $, and integration follows directly $\int \mathsf{a}(x) = \int a_\mu(x) \td x^\mu$. A vector is in this context called a 1-form. We can combine multiple 1-forms into $n$-form by using the wedge product $\wedge$, which is the antisymmetrized tensor product $\otimes$. Thus $\td x^\mu \wedge \td x^\nu =  \td x^\mu \otimes \td x^\nu - \td x^\nu \otimes \td x^\mu$. These definitions show that a 1-form describes a line element and a 2-form describes a surface element, and this language is therefore perfectly suited to describe our worldsheets.

Strictly speaking, the vortex worldsheet is a $d-2$-form and the current is a $d-1$ form (see Ref. \onlinecite{BeekmanZaanen11}). But we implicitly use the so-called Hodge duals of these quantities, turning them into $2$- and $1$-forms respectively, defined by $\mathsf{J}^\mathrm{V} = \mathcal{J}^\mathrm{V}_{\kappa\lambda}\;  \td x^\kappa \wedge \td x^\lambda$ and $\mathsf{w} = w_\mu \td x^\mu$. Based on the considerations laid out above, we propose that

{\em the vortex worldsheet $\mathcal{J}^\mathrm{V}_{\kappa\lambda}$  and the current $w_\mu = ( c \rho , \mathbf{w} )$ are related as
\begin{equation}\label{eq:vortex worldsheet current correpondence}
 \mathcal{J}^\mathrm{V}_{\kappa\lambda}\;  \td x^\kappa \wedge \td x^\lambda \sim \frac{1}{c\rho} w_\kappa \td x^\kappa \wedge w_\lambda \td x^\lambda.
\end{equation}
}
The factor $\frac{1}{c\rho}$ is necessary to obtain the correct dimensions, such that $\mathcal{J}^\mathrm{V}_{tl} = \frac{1}{c\rho} w_t w_l = w_l$ is a charge per time, agreeing with Eq. \eqref{eq:dimensionless 3+0D EoM}. In appendix \ref{sec:Transformation properties of current worldsheets} it is shown that this relation obeys the desired behavior under Lorentz transformations. We must conclude that, even though there is no closed form for the vortex equation of motion like there is for Abrikosov vortices Eq. \eqref{eq:Abrikosov vortex EoM}, the physical content is completely clear. Once the Mott condensate has formed, near the quantum phase transition it allows for vortex excitations which are lines of quantized current that obey Eq. \eqref{eq:vortex worldsheet current correpondence}.

Let us emphasize again that we have assumed here the strong type-II limit, corresponding to the London limit where fluctuations in the condensate amplitudes $|\Psi|$, $|\Phi|$ are suppressed. Phenomenologically this means that the penetration depth $\lambda$ is much larger than the coherence length $\xi$.  It is well understood how to interpret  the 
difference between type-I and type-II behavior in the duality context. In 2+1D (or 3D classical)\cite{HoveMoSudbo02} the type-I regime is associated with effective attractive interactions between the vortex particles, 
triggering a Van der Waals-type first order liquid--gas transition, translating into the first order transitions of the type-I state. The same logic of course applies to the 3+1D 
context where net attractions between vortex worldsheets would have a similar effect.  We just assume that the interactions are repulsive, prohibiting the clumping of vortex matter, such that the systems exhibit the continuous phase transition of the 3+1D/4D $XY$-model, equivalent to the type-II regime. 

\section{Phenomenology of the Bose-Mott insulator}\label{sec:Quantized current lines}
In this section we derive observable quantities of the Bose-Mott insulator and its vortices. The lack of a complete expression for the dual gauge fields is not as large an obstacle as one might think. The remainder of the paper will focus on the 3+0D limit of static vortex lines of stationary current, which will prove interesting enough. The complete dimensionful Lagrangian corresponding to Eqs. \eqref{eq:2+1D dual GL Lagrangian} and \eqref{eq:2+1D supercurrent Higgs Lagrangian} is,
\begin{align}\label{eq:Bose-Mott action}
\mathcal{L} &= \frac{a^{D-2}}{2J} (\epsilon_{mnk} \partial_n b_k)^2 + \frac{\hbar c_\text{ph}}{2 a^{D-1}} |\Phi|^2 (\partial_k \phi - \frac{a^{D-2}}{\hbar c_\textrm{ph}} b_k)^2 \nonumber \\
&\phantom{=} + \frac{e^*}{\hbar} (\epsilon_{mnk} \partial_n b_k) A_m - \frac{1}{4\mu_0} F_{mn}^2 - \frac{\tilde{a}}{2} | \Phi|^2 - \frac{\tilde{\beta}}{4}| \Phi|^4 .
\end{align}
Here $|\Phi|^2$ is the dimensionless condensate density. In what follows we will consider $D=3$ only.

We derive the equations of motions by varying Eq. (\ref{eq:Bose-Mott action}) with respect to $\bar{\Phi}$, $b_k$ and $A_m$.
The equations of motion are,
\begin{align}
 \frac{c\hbar}{a^2} \Phi -\tilde{\alpha}\Phi - \tilde{\beta} | \Phi|^2 \Phi &= 0,\label{eq:order parameter EoM}\\
  - \frac{a}{J} \epsilon_{knm} \partial_n w_m + \frac{| \Phi |^2}{a} (\partial_k \phi -  \frac{a}{\hbar c_\mathrm{ph}} b_k)   &= \frac{1}{2}\epsilon_{kmn} F_{mn} ,\label{eq:dual gauge field EoM}\\
  \frac{1}{\mu_0} \partial_n F_{nm} &= - \frac{e^*}{\hbar} w_m.\label{eq:electromagnetic EoM}
\end{align}
Here we have substituted the definition  $w_m= \epsilon_{mnk} \partial_n b_k$.  We are now ready to discuss the physical content of these equations. Note that the last two equations reduce to the equations of motion associated with the standard  Ginzburg--Landau superconductor in the limit $\vert \Phi \vert \to 0$.

\subsection{Maxwell equations}
The last equation Eq. (\ref{eq:electromagnetic EoM}) is clearly the inhomogeneous Maxwell equations for a source term $J^\mathrm{EM}_m = \frac{e^*}{\hbar} w_m$. These equations carry over from the superconductor, and do not pertain as such to the Mott insulating state. The insulating behavior is due to the expulsion of the electric current, which is represented by the term $\sim |\Phi|^2$. Therefore, Eq. (\ref{eq:electromagnetic EoM}) is just the vacuum contribution to dynamic electric and magnetic fields generated by a current source.

\subsection{Penetration depth}
The dual penetration depth $\lambda_\mathrm{M}$ sets the length scale up to which an electric current penetrates in the Mott insulating region. To identify it we act on Eq. (\ref{eq:dual gauge field EoM}) with $\epsilon_{rsk} \partial_s$. Contracting repeated indices, and using  $\partial_r w_r =0$, we find in the London limit of the dual superconductor $\vert \Phi \vert = \Phi_\infty$,
\begin{equation}\label{eq:dual penetration depth EoM}
 \frac{a}{J} \partial_m^2 w_r - \frac{\Phi_\infty^2}{\hbar c_\mathrm{ph} }  w_r +  \frac{e^*}{\hbar} \partial_n F_{nr} = - \frac{\Phi_\infty^2}{a} \mathcal{J}^\mathrm{V}_r.
\end{equation}
Here we recognize the Mott vortex current $\mathcal{J}^\mathrm{V}_r = \epsilon_{rsk} \partial_s \partial_k \phi$. The interpretation of this equation is as follows: a supercurrent $w_r$ can be generated by a vortex source $\mathcal{J}^\mathrm{V}_r$. This current is `dual Meissner screened' by the Mott condensate $\Phi_\infty$ as witnessed by the second term, but there is also some electromagnetic screening from the `backreaction' of the  induced electromagnetic field. In order to see this, we substitute Eq. (\ref{eq:electromagnetic EoM}) in this equation. In the absence of vortex sources, this leads to,
\begin{eqnarray}
 \frac{a}{J} \partial_m^2 J^\mathrm{EM}_r - \frac{\Phi_\infty^2}{\hbar c_\mathrm{ph}} J^\mathrm{EM}_r - \frac{\mu_0 {e^*}^2 }{\hbar^2} J^\mathrm{EM}_r &=& 0, \qquad 
 \textrm{or}\nonumber\\
\partial_m^2 J^\mathrm{EM}_r - \frac{\hbar \rho_\mathrm{s}}{c_\mathrm{ph} m^*} \Phi^2_\infty J^\mathrm{EM}_r - \frac{1}{\lambda_\mathrm{L}^2} J^\mathrm{EM}_r &=& 0.
\end{eqnarray}
Here we substituted $a/J = m^* / \hbar^2 \rho_\mathrm{s}$ [see Eqs. \eqref{eq:dimensionless coupling constants} and \eqref{eq:superfluid action}], and used the definition of the London penetration depth $\lambda_\mathrm{L}^2 = \mu_0 {e^*}^2 \rho_\mathrm{s} / m^*$. We find indeed two contributions to expulsion of electric current. The first $\sim \Phi_\infty^2$ is due to the Mott insulator, and the second remembers that the system originated from a superconductor. This is actually rather odd: the Meissner screening is due to the fact that the superconductor wants to expel the magnetic field, which is not true for the Mott insulator. Again one must consider that the insulator is a phase-disordered superconductor, and that on (very) short length scales the local boson superconductor is retrieved. Let us make a crude estimate of the relative strengths of the screening, by inserting the numerical values,
\begin{gather}
 \mu_0 = 4 \pi. 10^{-7} \approx 10^{-6} \mathrm{N}/\mathrm{A}^2, \quad e^* \approx 10^{-19} \mathrm{C},\nonumber\\
 \hbar \approx 10^{-34} \mathrm{J}\mathrm{s}, \quad c_\mathrm{ph} \approx \frac{1}{300} c \approx 10^6 \mathrm{m}/\mathrm{s},
\end{gather}
we find that the relative strengths are,
\begin{equation}\label{eq:Mott/Meissner screening ratio}
 \frac{ \mathrm{Mott} }{\mathrm{Meissner}} \approx \frac{ \Phi_\infty^2}{ \mu_0 {e^*}^2 c_\mathrm{ph} / \hbar } \approx 10^4 \Phi_\infty^2.
\end{equation}
Here $\Phi_\infty^2$ is dimensionless, but as the order parameter of the Mott condensate it should be large. Therefore the expulsion of electric current due to the Mott term is several orders of magnitude stronger than the Meissner screening, and for all purposes the latter may be ignored, also eliminating our interpretation problem. 

Hence the ``Mott proximity depth'' for electric current is $\lambda_\mathrm{M} = \sqrt{ \frac{\hbar}{c_\mathrm{ph} m^*} \rho_\mathrm{s} \Phi_\infty^2}$. It depends on a number of material parameters. We encounter the ubiquitous combination $\rho_s \Phi_\infty^2$, which is the product of the superconducting order parameter and the Mott order parameter. At first, one may think that they should be mutually exclusive, as one has either superconducting order \emph{or} Mott insulating order. However one must realize that the Mott insulator is formed from repelling Cooper pairs: the larger the number of Cooper pairs, as quantified by the superfluid density $\rho_\mathrm{s}$, the stronger the electromagnetic effects such as screening will be. It is just $\Phi_\infty^2$ that signals the existence of the Mott state, whereas the combination $\rho_\mathrm{s}\Phi_\infty^2$ is the appropriate dual Higgs mass of this insulator.

\subsection{Coherence length}
Upon rescaling the dual order parameter $\Phi$ in Eq. \eqref{eq:order parameter EoM} by dividing it by its equilibrium value $\Phi_\infty = \sqrt{\frac{|\tilde{\alpha}|}{\tilde{\beta}}}$, such that $\Phi = \Phi_\infty \Phi'$, and taking $b_k=0$ as is the case deep within the Mott insulator, this equation reduces to,
\begin{equation}
 \frac{a^2}{| \tilde{\alpha}|} (\partial_m)^2 \Phi' + \Phi'- \Phi^{\prime 3} =0.
\end{equation}
This shows that we can define a dual coherence length as $\tilde{\xi} = \frac{a}{\sqrt{| \tilde{\alpha} }}$, depending on the details of the symmetry breaking through the precise value of the Ginzburg--Landau parameter $| \tilde{\alpha}|$. Fluctuations in the order parameter will take place within a typical length $\tilde{\xi}$. In the strong type-II limit $\lambda_\mathrm{M} > \tilde{\xi}$.

\subsection{Current quantization}
We are now in the position to present our central result: the quantization of vortex lines of electric supercurrent. Eq.  (\ref{eq:dual gauge field EoM}) is similar to the regular Ginzburg--Landau equation. Consider a closed contour over which the change of the phase $\phi$ is a multiple of $2\pi$,
\begin{equation}\label{eq:current quantization line integral}
 \oint_{\partial \mathcal{S}} \td x^k\ \partial_k \phi = 2\pi N.
\end{equation}
We are free to choose this contour deep within the Mott insulator far away from the vortex line, such that the electric current is suppressed $w_m =0$. Assume that  there is no external electromagnetic field $F^\mathrm{ext}_{mn} = 0$, while the induced field is very small as argued in Eq. \eqref{eq:Mott/Meissner screening ratio}. Under these conditions Eq. \eqref{eq:dual gauge field EoM} reduces to,
\begin{equation}
\frac{\hbar c_\mathrm{ph}}{a} \partial_k \phi  = b_k.
\end{equation}
Taking the line integral of this equation as in Eq. \eqref{eq:current quantization line integral}, invoking  Stokes' theorem on the right-hand side, one finds,
\begin{align}
\frac{\hbar c_\mathrm{ph}}{a} 2\pi N 
  &= \frac{\hbar c_\mathrm{ph}}{a} \oint_{\partial {\cal S}} \td x^k\ \partial_k \phi =\oint_{\partial {\cal S}} \td x^k\ b_k\nonumber\\
  & = \int_{\cal{S}} \td S_m\ \epsilon_{mnl} \partial_n b_{l} = \int_{\cal{S}} \td S_m\ w_m.
\end{align}
The right-hand side is the flux of current $w_m$ through the surface $\mathcal{S}$. Since the current is expelled from the Mott insulator, this current flows through the vortex line. For the electric current $I$ which is the flux of the current density $J_m = \frac{e^*}{\hbar} w_m$, this implies the quantization condition,
\begin{equation}\label{eq:current quantum}
 I_0 = \frac{e^*}{\hbar} \frac{\hbar c_\mathrm{ph}}{a} 2\pi N = \frac{1}{\Phi_0} \sqrt{UJ} (2 \pi)^2 N.
\end{equation}
Here $\Phi_0 = h/e^*$ is the (magnetic) flux quantum and we have substituted the microscopic parameters $\sqrt{UJ} = \hbar c_\mathrm{ph} /a $ from Sec. \ref{sec:Bose-Hubbard model}. Following similar arguments as for the Abrikosov lattice, higher winding number ($N>1$) vortices are energetically unfavorable, and split up into multiple $N=1$ lines. Each such vortex line carries an electric current,  for typical values $c_\mathrm{ph} \approx 10^6\; \mathrm{m}/\mathrm{s}$ and $a \approx 10^{-10}\; \mathrm{m}$, of,
\begin{equation}\label{eq:current quantum estimate}
 I_0 = \frac{e^* c_\mathrm{ph}}{a} 2\pi \approx 10^{-2} \mathrm{A}.
\end{equation}

Different from the magnetic flux quantum, this current quantum is not exclusively expressed in fundamental constants for the obvious reason that the unit of current (Amp\`ere) cannot be composed this way. However, we do observe that the current quantum is inversely proportional to the flux quantum in Eq. \eqref{eq:current quantum} as expected by duality. However, one needs a quantity with dimension of energy ($\sqrt{UJ}$) to convert the reciprocal of the flux quantum into a quantum carrying the dimension of charge per time. This then conspires into the combination of superconducting phase velocity and lattice constant of Eq. \eqref{eq:current quantum}. Although depending on material specifics, $ c_\mathrm{ph} /a$ is expected to be a fixed quantity in the proximity of the superconductor--Mott insulator quantum phase transition.

\section{Phase diagram}\label{sec:Phase diagram}
\begin{figure*}
\begin{center}
 \includegraphics[width=12cm]{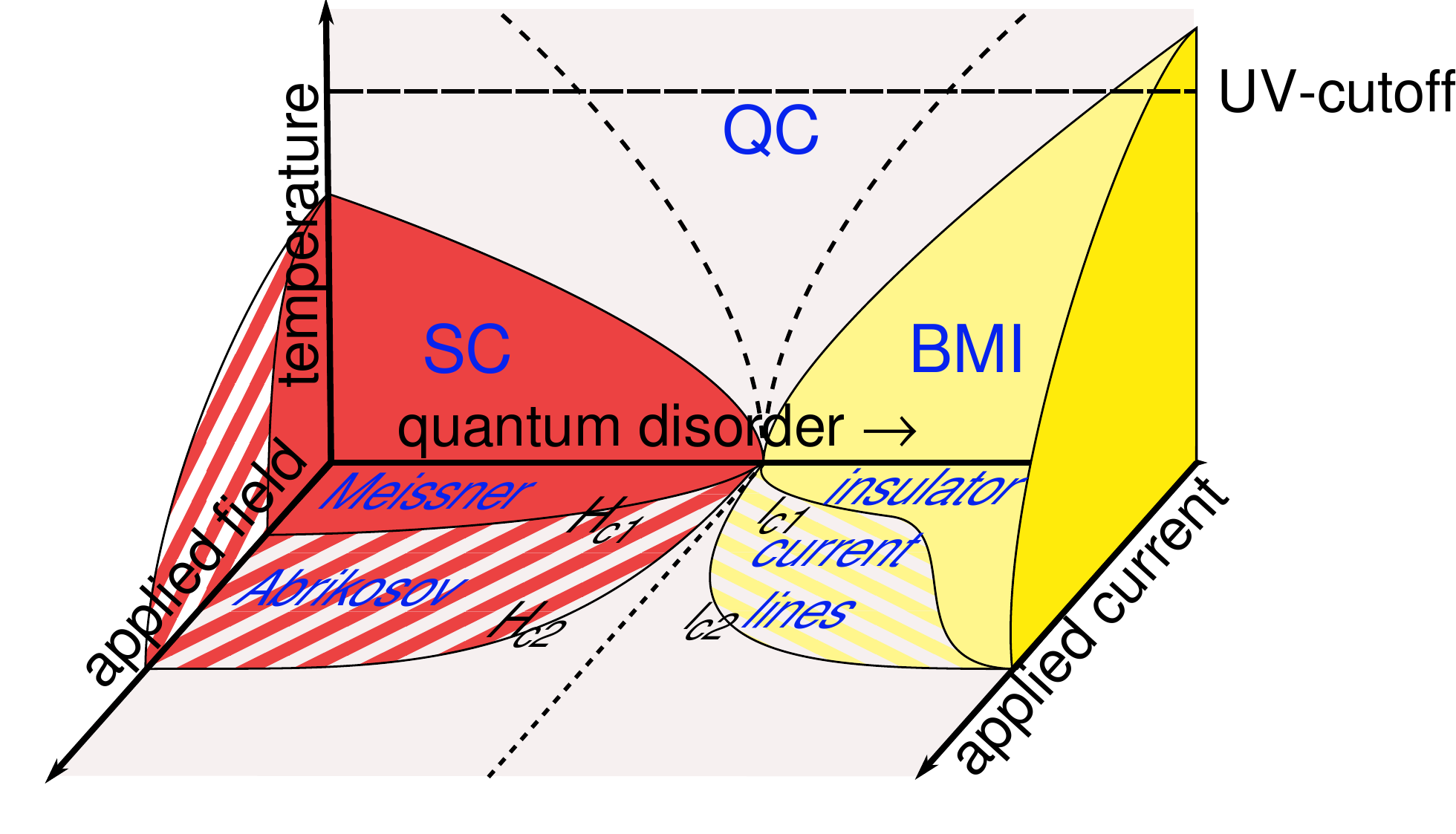}
\caption{Topology of the phase diagram of type-II Bose-Mott insulators in three space dimensions. The horizontal axis represents strength of quantum phase fluctuations that tune from superconducting (SC) to  Bose-Mott insulating (BMI) order via a quantum phase transition. As function of temperature (vertical axis) 
the ``wedge'' of quantum critical fluid (QC) emerges from the quantum critical point, while on either side thermal phase transitions occur: obviously to a
 superconducting state,  but duality also predicts a thermal transition on the insulating side. 
 An external magnetic field is first expelled by the superconductor (Meissner), while above $H_{\mathrm{c1}}$ it penetrates as an Abrikosov  lattice of quantized flux lines.
 Near the quantum phase transition a similar response is observed in the Mott insulator, now as function of applied {\em electric 
current}. First the current is expelled, just an insulator. However, above a `lower critical current'  $I_{\mathrm{c1}}$  current will penetrate 
as an `Abrikosov lattice' of lines carrying a quantized current flux: the type-II Mott insulator. Upon increasing the local repulsions one eventually enters the regime of nearly completely localized bosons. Here the type-II Mott insulator disappears
because the relevant length scales have shrunk beyond the lattice constant: the reason the type-II phase has hitherto been overlooked.} \label{fig:Mott phase diagram}
 \end{center}
\end{figure*}

The results in the above lead us to propose a general phase diagram for the charged Bose-Hubbard model in 3+1D (Fig. \ref{fig:Mott phase diagram}). The control parameters are i) the (quantum) coupling constant $g \sim \sqrt{U/J}$; ii) temperature; and iii) an applied magnetic field for the superconductor, or an applied electric current for the Bose-Mott insulator.

The superconducting side is well known. At $U \to 0$ we have the familiar $H$--$T$-diagram for type-II superconductors (we always assume local pairs, $\xi \ll \lambda_\mathrm{L}$). At low applied field the Meissner phase is formed with complete field expulsion. Above $H_\mathrm{c1} \sim \Phi_0$ an Abrikosov vortex lattice forms, until the superconductivity is destroyed at $H_\mathrm{c2}$. 

Upon approaching the superconductor--Mott insulator transition by increasing the coupling constant $g$, the renormalized phase stiffness is decreasing and the superconducting $T_\mathrm{c}$ vanishes right at the quantum phase transition.

Contrary to standard expectation that on the Mott insulating side of this quantum phase transition a featureless state is found, the first surprising ramification of our duality is that in 3+1D there is actually a thermal phase transition in the Mott state with a finite critical temperature that is rising (at zero applied current) upon moving away from the QPT. As usual, the thermal phase transition occurs on both sides of the QPT well inside the renormalized classical regime; upon raising temperature one will re-enter the quantum critical `wedge'.

Duality dictates that the magnetic field in the superconductor turns into electric current for the Bose-Mott insulator. Hence, the control parameter equivalent to the magnetic field of the superconductor is applied current for the Bose-Mott insulating side. Near the quantum phase transition where duality is expected to be universally valid, this suggests a dual Meissner phase (insulator) below the current quantum $I_0$, while at higher current a dual Abrikosov lattice of quantized current lines will occur. When current gets too high, above the upper critical current $I_{c2}$, the Bose-Mott insulator will be completely destroyed. The resulting state is again a superconductor, since the dual of the Bose-Mott insulator is a superconductor and not a metal. Of course, if the applied current exceeds the pair-breaking current density of the superconductor, superconductivity is destroyed, and the Bose-Hubbard model no longer makes sense.

Upon increasing $U/J$ further, eventually one ends up in the simple strong-coupling limit of the Bose-Mott insulator, with the nearly completely localized bosons subjected to short range virtual fluctuations. This is indeed a featureless state, one that sets the common intuition. How to accommodate the rich physics we claim near the quantum phase transition? This is subtle, but in fact quite simple. In the language of the dual superconductor, the dual penetration depth shrinks to the lattice constant: the effect is that our current-carrying vortices `fall through the lattice constant'. Their core energy exceeds the UV cut-off and the the ``type-I Bose-Mott insulator'' can be formed, with a $T_\mathrm{c}$ that has `disappeared above the cut-off'. In this limit the state turns featureless again, just governed by the thermal excitations of the `massive photons' (doublons/holons) of the dual Higgs condensate.

The strong coupling is therefore part of the reason for the misleading intuition that the Bose-Mott insulator has to be thermodynamically featureless. Another flaw in this regard is associated with approaching it from the other side: it is well understood that the quantum phase transition in 3+1D is at the upper critical dimension, as it is effectively $XY$ in 4D. The critical regime is therefore governed by mean-field and dominated by amplitude fluctuations. How then can these vortices play such a central role?

This is just a confusion based on overestimation of the influence of universality class away from the critical regime. A central result of the renormalization group/critical theory is that the ``soft spin'' ($\varphi^4$) and ``hard-spin'' (sigma model, our Bose-Hubbard model) share the same, universal critical regime which is surely of the mean-field kind. However, duality is actually relating the (stable) fixed points on both sides of the phase transition: our dual superconductor becomes discernible only well below the cross-over to the quantum critical regime. The quantum critical regime is itself governed by the mean-field dynamics of 4D $XY$.

Surely, starting with microscopic circumstances that directly coarse grain in a $\varphi^4$-theory, there will be no interesting physics on the insulating side: this physics plays no role in a real BCS superconductor (such as aluminum) since amplitude fluctuations dominate on all scales. However, starting out with strongly bound, hard-core bosons, duality cannot be avoided. All that matters is that the dual penetration depth be large compared to the lattice constant near the quantum phase transition. This length scale is coincident with the typical distance between free vortices and as long as this is large compared to the lattice constant the Mott insulator must be the dual vortex superconductor. With regard to the critical regime one just learns that the proliferation of the vortex strings is described by a mean-field regime, regardless of whether these strings carry magnetic fluxes (superconducting side) or electric currents (Mott side): it remains an interesting exercise to find 
out why these critical theories match in one common critical regime\cite{NguyenSudbo99,HoveSudbo00, Kleinert08}.

Summarizing, the type-II Bose-Mott insulator is a dual type-II superconductor. The dual of the magnetic field of the superconductor, is electric current for the Bose-Mott insulator. To probe a superconductor, one applies a magnetic field from outside. Outside the superconductor there is a medium which supports a magnetic field with magnetic permeability $\mu$. It does not matter whether this is a vacuum with permeability $\mu_0$ or a dielectric with another value. Even a metal would do when considering a static magnetic field. As such, there are only two inequivalent phases with regard to magnetic field: Meissner and not-Meissner. 

For electric current, the situation is different. The superconductor is to the Bose-Mott insulator what the (Maxwell) vacuum is to the superconductor. The superconductor carries the current applied from outside to the Bose-Mott insulator. This begs following the question: What is the Maxwell vacuum to the Bose-Mott insulator? The vacuum has no charge carriers, and does not support a current. Also a metal is different, since a dissipative current of quasiparticles (fermions) may be different from a supercurrent made out of Cooper pairs. Therefore we anticipate that the situation for the type-II Bose-Mott insulator is richer than that for the superconductor. This was already mentioned in Table \ref{table:SC MI correspondences}, and is also exploited in the experimental setups of the next section.

\section{Proposed experiments}\label{sec:Proposed experiments}
It is not an accident that this type-II Bose-Mott insulator was never seen in the laboratory. The best model systems are either hard to realize in three dimensions (Josephson 
networks) or it is unclear how to impose external currents (cold atoms), while in conventional condensed matter systems it is uncertain whether such physics is at work at all.  In general one should focus on systems with large phase fluctuations. Obvious candidates are strongly underdoped  cuprate 
superconductors. Here the elusive pseudogap phase is by many conjectured to consist of so-called preformed Cooper pairs, which are bosons. Thus at some high temperature $T^*$ the pairs bind into bosons, and only at some lower temperature $T_\mathrm{c}$ phase coherence sets in, leading to superconductivity\cite{EmeryKivelson95a,HartnollKovtunMullerSachdev07,FranzTesanovic01}. Hence, the quantum phase transition from superconductor to Bose-Mott insulator would be precisely of the vortex-proliferation kind discussed here. Experimental support comes from the Nernst effect  \cite{WangLiOng06}, diamagnetic behavior \cite{LiEtAl10} and spectroscopy \cite{LeeEtAl09}. One can also wonder whether the ``giant proximity effect'' associated with a 100nm thick underdoped cuprate 
barrier layer \cite{BozovicEtAl04,MorenzinoEtAl11} has dealings with type-II Bose-Mott behavior (see Appendix \ref{sec:Giant proximity effect}). If the pseudogap indeed consists of phase-incoherent local bosons, the type-II Bose-Mott insulator should be found close to the quantum phase transition and at low temperatures. Conversely, if the current line lattice is found in the underdoped cuprates, it would constitute indirect but convincing evidence of the existence of preformed Cooper pairs. Another candidate may be the so-called disordered or amorphous superconductors, see for instance Ref. \onlinecite{SacepeEtAl11}.

\begin{figure*}
\begin{center}
\includegraphics[width=12cm]{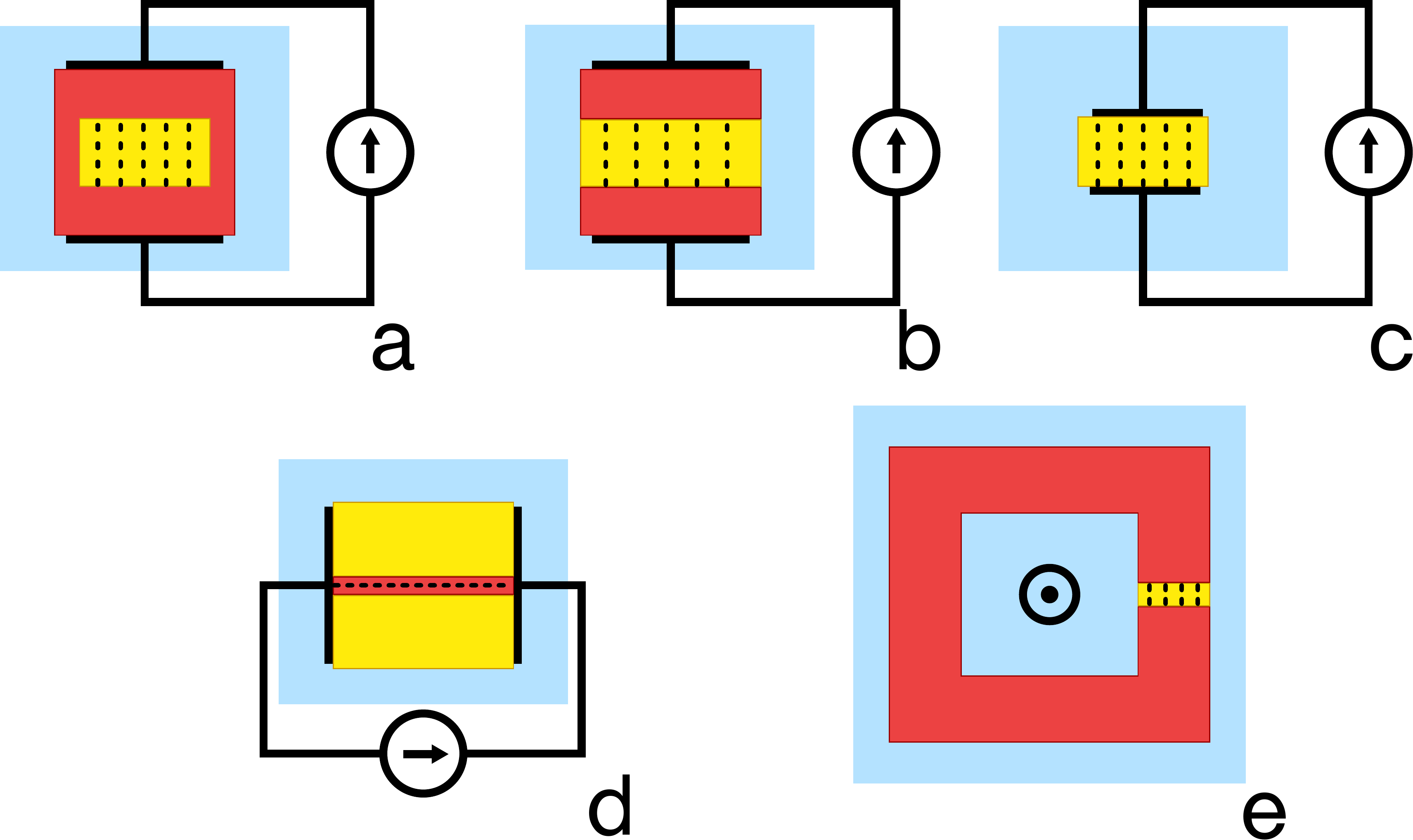}
\caption{Various devices to measure the type-II Bose-Mott insulator. Indicated are the type-II Bose-Mott insulator (MI) in yellow, the superconductor in red and the Maxwell vacuum in blue, with the current lines as dashed black lines. The circle and arrow represent a current source. 
(a) The precise dual of the type-II superconductor: the outside superconductor carrying current acts like the Maxwell vacuum carrying a magnetic field, which is then 
penetrating the type-II MI in the form of the quantized flux lines. A complication is that the currents in the outside superconductor decay over the London penetration depth. (b) A Josephson junction using the MI as barrier. 
For a very small bias the barrier will be insulating, to become completely transparent above  
$I_{\mathrm{c1}}$. (c) The medium imposing the current need not be a superconductor: a metal works as well. A simple ``capacitor'' will short circuit at $I_{\mathrm{c1}}$.  (d) Equivalent of Josephson vortices where the quantized vortex line does not form inside MI but within a narrow junction layer of SC. (e)  SQUID setup in which current bias is increased in very small steps by a perpendicular magnetic field (circle with dot). Current will not flow until the first vortex is formed.
}\label{fig:Mott vortex experiments}
\end{center}
\end{figure*}

Even though the imagination of experimentalists would be more qualified to devise the most suitable setup, we wish here to sketch some ideas to probe the current line lattice. In Fig. \ref{fig:Mott vortex experiments} we illustrate a number of possible devices to measure the type-II Bose-Mott insulator, all revolving around the basic ingredient of imposing an external current on the Mott insulator.

Setups (a)-(c) involve the classic junction type. One tries to force current through a slab of type-II Bose-Mott insulator. The first experiment (a) is analogous to the type-II superconductor. There, magnetic fields lines penetrate from the outside to form the first Abrikosov vortex. The current at first flows around the type-II Bose-Mott insulator, and when the current is large enough (higher than the current quantum $I_0$) it will `trickle in' from the outside to form the first current line. The thickness of the superconductor surrounding the type-II Bose-Mott insulator should be smaller than the penetration depth $\lambda_\mathrm{L}$, since in a superconductor current flows near the edges. Experiment (b) is similar, but there are no superconducting walls. This presumes that current lines will form internally, not coming in from the outside. Below the lower critical current $I_\mathrm{c1}$ no current will flow at all, making for a stronger signal. Experiment (c) has normal leads instead of superconductors; 
nevertheless possibly any applied current suffices to create vortex lines of supercurrent. In these experiments, the slab of type-II Bose-Mott insulator should be thick, to preclude any normal Josephson effect. In all these cases, the measured signal would be a sudden jump in the current when a current vortex line is formed.

Experiment (d) connects with so-called Josephson vortices in superconductors. In a Josephson junction, a narrow barrier in between two superconductors, vortices of quantized magnetic field \emph{along the junction} can form under applied field, which behave like Abrikosov vortices except that they do not have a normal core. Here a small barrier of superconductor is sandwiched in between two layers of type-II Bose-Mott insulator. The current through the superconductor would be quantized if the dual phase coherence of the type-II Bose-Mott insulator imposes on the narrow barrier.

At first sight the SQUID setup (e) looks particularly promising. One would like to impose a current bias, and in the junction-type experiments above, one actually applies a potential bias. One of the consequences could be that, since in general (Bose-)Mott insulators are rather poor insulators, leak currents may spoil the signal. In setup (e) a ring of superconductor is interrupted by a thick layer of type-II Bose-Mott insulator; this is a typical superconducting quantum interference device (SQUID), except that the barrier is intentionally very thick to preclude the normal Josephson effect. Applying a magnetic field through the SQUID loop will cause a phase difference across the insulating barrier. This phase difference will not induce a normal dissipative current, but if this difference is large enough, a vortex line of supercurrent may form. Then a current will flow through the loop with magnitude of one current quantum. Measuring the magnetic field through the loop (for instance with a second SQUID), one 
would see a sudden drop when this current starts to flow. Increasing applied magnetic further would induce more and more current lines. The current quantum in a high-$T_\mathrm{c}$ material
is estimated to be quite large  [$I_0 \approx 10^{-2} \mathrm{A}$, see Eq. \eqref{eq:current quantum estimate}]. However, we would not be surprised when the pinning of these current lines would turn 
out to be very strong, given for instance the strong spatial inhomogeneity  of the superconducting order in this regime as observed by scanning tunneling 
spectroscopy \cite{LangEtAl02}.
This might cause substantial ``current flux penetration'' difficulties, in analogy with the complications that are well documented in the context of the usual vortex
dynamics \cite{BlatterEtAl94,RosensteinLi10}. As for the Abrikosov lattice, the most
direct way to probe the type-II Bose-Mott insulator would be the analog of decoration experiments, directly imaging the current lines. Scanning tunneling spectroscopy is here
an option with the caveat that the size of the current line is set by the Mott proximity depth which can be quite large near the QPT. Alternatives are microwave impedance 
or low energy electron microscopy measurements. 

\section{Conclusions}\label{sec:Conclusions}
We have demonstrated that Bose-Mott insulators which are close to the quantum phase transition to the superconductor exhibit a much richer physics than 
the intuition that follows from the strong-coupling limit would indicate. According to a precise quantum field-theoretical duality, its physics should be a mirror image of the rich physics of
normal superconductors. The highlight is our prediction of the existence of an analog of the type-II phase, where now an Abrikosov lattice is formed of topological vortex lines
that carry a quantized flux of supercurrent. Since these Mott insulators have to be three dimensional, while the type-II phase is induced by currents that are imposed from 
the outside, it is not straightforward to see these effects in model systems that are designed to represent the Bose-Hubbard problem in the laboratory (cold atoms, Josephson networks). However, the type-II effect can be exploited to find out whether such physics is indeed at work in underdoped high-$T_\mathrm{c}$ superconductors. It is imaginable that there is a range in dopings near
the quantum phase transition where the low-$T_\mathrm{c}$ and low superfluid density superconductors will turn out to be Mott insulators camouflaged as superconductors due to a glassy
current line network induced by the measurement fields.

\acknowledgments{
We thank J. Aarts, J.C. Davis, I.F. Herbut, H. Hilgenkamp, P.H. Kes, Z. Te\v{s}anovi\'{c}  for useful discussions and especially J.M. van Ruitenbeek for suggesting the setup in Fig. \ref{fig:Mott vortex experiments}(e). This work was supported by the Netherlands foundation for Fundamental Research of Matter (FOM) and the Nederlandse Organisatie voor Wetenschappelijk Onderzoek (NWO) via a Spinoza grant. A.J.B. is also supported by the Foreign Postdoctoral Researcher program at RIKEN.
}

\appendix
\section{Giant proximity effect}\label{sec:Giant proximity effect}
In the regular Josephson effect, a supercurrent can flow between two superconductors, even if there is a spatial gap or barrier in between them. Because the superconducting order parameters extend outside the superconductor, if the barrier is narrow enough that the two order parameters overlap, the supercurrent is supported even within the barrier. The order parameters fall off exponentially with typical tunneling length scale that is microscopic. If the region in between is a good metal, instead the normal metal coherence length $\xi_\mathrm{n}$ is the appropriate length scale.

\begin{figure}
\begin{center}
  \includegraphics[width=4cm]{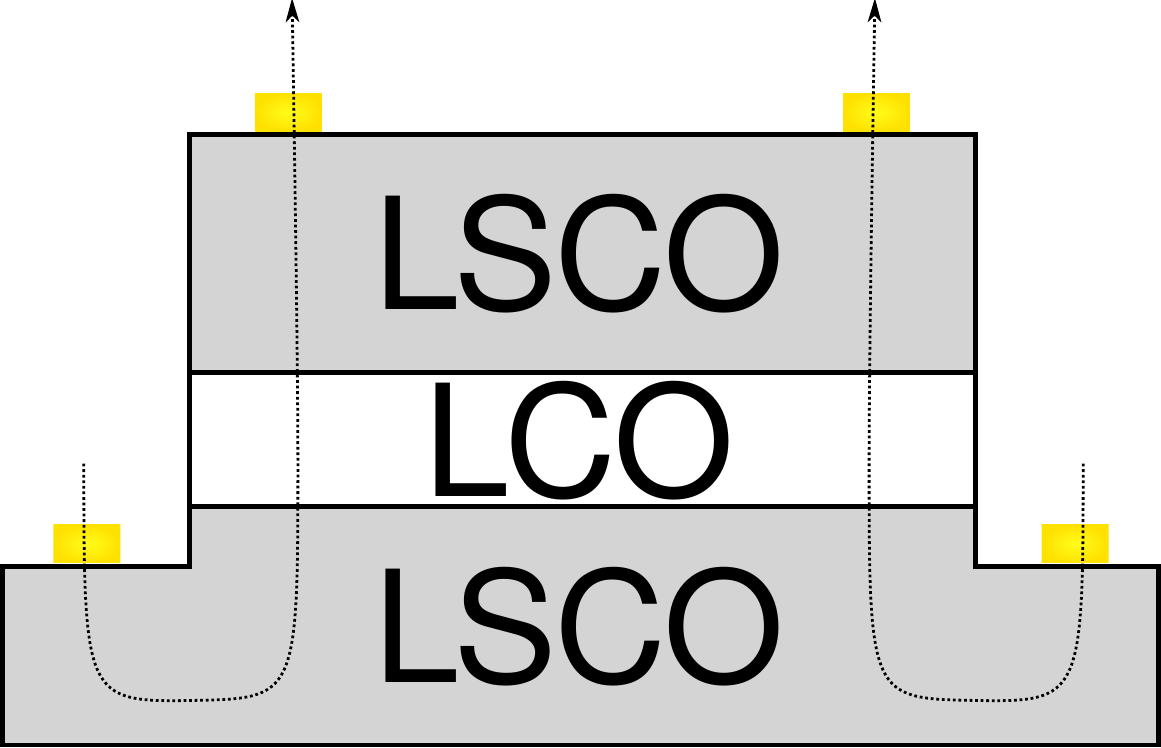}
 
\end{center}
 \caption{The geometry of the experiment of Ref. \onlinecite{BozovicEtAl04} on the giant proximity effect. Between two layers of  La$_{1.85}$Sr$_{0.15}$CuO$_4$ ($T_\mathrm{c} = 40$K) is one layer of La$_2$CuO$_{4+\delta}$ ($T_\mathrm{c} = 20$K) that can be up to more than 100 \AA{}  thick. The yellow spots indicate gold leads over which a voltage bias can be installed. At $T = 30$K supercurrent was observed to flow (indicated by the dashed profiles) even through these thick barriers that greatly exceed the coherence length of only  several \AA.
In Ref. \onlinecite{MorenzinoEtAl11} a similar geometry was used with leads of La$_{1.84}$Sr$_{0.15}$CuO$_4$ ($T_\mathrm{c} \approx 32$K) and the layer in between of La$_{1.94}$Sr$_{0.06}$CuO$_4$ ($T_\mathrm{c} < 5$K) that is 46nm thick. Muon spin resonance measurements at decreasing temperatures showed an increasing diamagnetic response to an applied field of 9.5 mT.}\label{fig:GPE}
\end{figure}

Now, in type-II superconductors and especially in the high-$T_\mathrm{c}$ cuprates, the tunneling length is very short (of the order of several \AA{}). It came therefore as a surprise that placing an underdoped, non-superconducting cuprate layer between two cuprate superconducting leads (see Fig. \ref{fig:GPE}), a current was observed to flow even if the barrier was very wide, up to hundreds of \AA{}f. A review of this ``giant proximity effect'' can be found in Ref. \onlinecite{DelinKleinsasser96}. It was speculated that the superconductor would Josephson couple between impurities throughout the barriers, but the very clean samples of Ref. \onlinecite{BozovicEtAl04} seemed to be the definitive answer that really a new phenomenon comes into play. Several explanations were proposed \cite{KresinOvchinnikovWolf03,AlvarezMayrMoreoDagotto05,CovaciMarsiglio06,MarchandCovaciBerciuFranz08}, all of which have in common that superconducting order is induced homogeneously within the barrier region.

When the barrier region is a type-II Bose-Mott insulator as proposed in this work, the giant proximity effect will be established in a qualitatively very different manner. The supercurrent would penetrate as vortex lines of electric current, leading to very inhomogeneous conductivity as the vortices form a current line lattice. This explains very simply how the Josephson effect is ``giant'': the energy cost of the formation of a vortex line is linear in its length, while the regular Josephson effect is limited by the exponential fall-off of the order parameter. A clear test to confirm this prediction would be to observe the spatial variation of the conductivity once the supercurrent is flowing.

On a related note, very recently it was observed in sandwich samples very similar to those considered in the previous paragraphs, that the barrier region through which (super)current flows undergoes a Meissner effect: applied magnetic field perpendicular to the $c$-axis of the junction, is expelled \cite{MorenzinoEtAl11} by a diamagnetic response. Therefore we may regard this region to be superconducting. This agrees nicely with our proposed phase diagram, where an applied current above the higher critical current $I_{c2}$ destroys the Mott insulator and at low enough temperatures drives the the system back to the superconducting state. The applied field would induce a countercurrent in the superconducting leads that causes the type-II Bose-Mott insulator in between to permit current lines, or is even pushed above its upper critical current $I_{c2}$ to become completely superconducting. The authors of this work claim that electric current must also be flowing in the $ab$-plane in which the cuprate layer lies, in addition to flow across the junction, which would surely favor the latter scenario.

\section{Transformation properties of current worldsheets}\label{sec:Transformation properties of current worldsheets}
Here we show that the relation between the vortex worldsheet and the electric current Eq. \eqref{eq:vortex worldsheet current correpondence},
\begin{equation}\label{eq:vortex worldsheet current correpondence repeat}
 \mathcal{J}^\mathrm{V}_{\kappa\lambda}\;  \td x^\kappa \wedge \td x^\lambda \sim \frac{1}{c\rho} w_\kappa \td x^\kappa \wedge w_\lambda \td x^\lambda,
\end{equation}
has the correct properties under Lorentz transformations. That is, the worldsheet $J^\mathrm{V}_{\kappa\lambda}$ is a Lorentz tensor and the electric current $w_\mu$ is a Lorentz vector.  

Start out from a static vortex line in the $z$-direction. The only non-zero component of the vortex worldsheet is $\mathcal{J}^\mathrm{V}_{tz}$ and the current is $w_\mu = (c\rho , 0 , 0, w_z)$. Perform a rotation in the $xz$-plane over angle $\alpha$. The transformed fields are,
\begin{align}
 \begin{pmatrix}
  J'_{tx} \\ J'_{tz} \\ J_{tx}
 \end{pmatrix}
=
\begin{pmatrix}
  \sin \alpha \; J_{tz}  \\ \cos \alpha\;  J_{tz} \\ 0 
 \end{pmatrix} \\
 \begin{pmatrix}
  \rho' \\ w'_x \\ w'_z 
 \end{pmatrix}
 =
 \begin{pmatrix}
  c\rho \\ \sin \alpha \; w_z \\ \cos \alpha \; w_z
 \end{pmatrix}\label{eq:rotated current}
\end{align}
It is easily verified that relation Eq. \eqref{eq:vortex worldsheet current correpondence repeat} holds.

Next, from the static line in the $z$-direction, perform a Lorentz boost in the $x$-direction with velocity $v$. The Lorentz transformation matrix is,
\begin{equation}
 \Lambda_\mathrm{boost} = 
 \begin{pmatrix}
  \gamma &\gamma \beta& & \\
  \gamma \beta & \gamma & &  \\
  & & 1 & \\
  &  & & 1
\end{pmatrix}.
\end{equation}
Here $\gamma = \frac{1}{\sqrt{1-v^2/c^2}}$ and $\beta = v/c$. The transformed fields are,
\begin{align}
 \begin{pmatrix}
  J`_{tx} \\ J`_{tz} \\  J`_{xz}
 \end{pmatrix}
=
\begin{pmatrix}
  0 \\ \gamma J_{tz} \\ \gamma\beta  J_{tz} 
 \end{pmatrix} \\
 \begin{pmatrix}
  c\rho` \\ w`_x \\ w`_z 
 \end{pmatrix}
 = 
 \begin{pmatrix}
  \gamma c\rho \\ \gamma \beta c\rho \\   w_z
 \end{pmatrix} \label{eq:boosted current}
\end{align}
 The worldsheet is no longer pointing `straight up' in the time direction, but is tilted with a non-zero component $J`_{xz}$, denoting the motion in the $x$-direction of the vortex line along $z$. For an Abrikosov vortex this component would correspond one-to-one with an electric field in the $y$-direction. But as we see, it now leads to a component of the electric current in the $x$ direction. If we compare the two currents Eqs. \eqref{eq:rotated current} and \eqref{eq:boosted current}, then given a 4-current $w_\mu$, we cannot say whether it corresponds to a stationary or  a moving vortex line. We need additional information, such as time derivatives. Relation Eq. \eqref{eq:vortex worldsheet current correpondence repeat} holds because $\frac{1}{c\rho} |w`_x \wedge w`_z| = \frac{1}{c\rho}\gamma \beta c\rho w_z = \gamma \beta w_z = | J`_{xz} |$.
 
 Now look at a Lorentz boost of the rotated current, which already has a component in the $x$-direction. The transformed fields are,
 \begin{align}
 \begin{pmatrix}
  J''_{tx} \\ J''_{tz} \\  J''_{xz}
 \end{pmatrix}
=\Lambda_\mathrm{boost} \rightharpoonup
\begin{pmatrix}
 \sin \alpha \; J'_{tz}  \\ \cos \alpha\;  J'_{tz} \\ 0 
 \end{pmatrix}
 =
 \begin{pmatrix}
 \sin \alpha \; J_{tz}  \\ \gamma \cos \alpha\;  J_{tz} \\ \gamma \beta   \cos \alpha\;  J_{tz}
 \end{pmatrix}\label{eq:rotated and boosted worldsheet}
 \\
 \begin{pmatrix}
  c\rho'' \\ w''_x \\ w''_z 
 \end{pmatrix}
 =\Lambda_\mathrm{boost} \rightharpoonup
 \begin{pmatrix}
 c\rho' \\ w'_x \\   w'_z
 \end{pmatrix} 
 =
 \begin{pmatrix}
 \gamma c\rho + \gamma\beta \sin \alpha \; w_z \\ \gamma\beta c\rho + \gamma \sin \alpha \; w_z \\   \cos \alpha \; w_z
 \end{pmatrix} 
 \label{eq:rotated and boosted current}
\end{align}
For the boost of $J'_{tx}$, this follows from $J''_{tx} = \Lambda_t^{\phantom{t}t} \Lambda_x^{\phantom{x}x} J'_{tx} +  \Lambda_t^{\phantom{t}x} \Lambda_x^{\phantom{x}t} J'_{xt} = \gamma^2  J'_{tx} + \gamma^2 \beta^2 J'_{xt} = (\gamma^2  - \gamma^2 \beta^2)J'_{tx} = J'_{tx}$. This component is invariant under boosts in the $x$-direction, because the contraction in the spatial $x$-direction is `compensated' by the dilation in the temporal direction. The surface area of elements in the $xt$-plane is therefore unchanged. For the corresponding current component this follows from,
\begin{align}
 w''_t &\td t'' \wedge  w''_x \td x'' \nonumber\\ 
 &= ( \gamma w'_t \td t' + \gamma \beta w'_x \td x' ) \wedge (  \beta \gamma w'_t \td t' + \gamma  w'_x \td x' ) \nonumber \\
 &= \gamma^2 w'_t w'_x \td t' \wedge \td x' + \gamma^2 \beta^2 \td x' \wedge \td t' \nonumber\\
 &= (\gamma^2 - \gamma^2 \beta^2) w'_t w'_x \td t' \wedge \td x' \nonumber \\
 &=  w'_t \td t' \wedge w'_x \td x'.
 \end{align}
Here we used the antisymmetry of the wedge product $\wedge$. For the verification of the $xz$-component we compute,
\begin{align}
 w''_x \td x'' \wedge  w''_x \td x''
 &=  (  \beta \gamma w'_t \td t' + \gamma  w'_x \td x' ) \wedge w'_z \td z' \nonumber \\
 &= (  \beta \gamma c \rho \td t + \gamma \sin \alpha \; w_z \td z)  \wedge \cos \alpha w_z \td z \nonumber \\
 &= \beta \gamma c \rho \cos \alpha w_z  \td t \wedge \td z \nonumber\\
 &=   \beta \gamma \cos \alpha J_{tz}  \td t \wedge \td z.
\end{align}
This agrees with Eq. \eqref{eq:rotated and boosted worldsheet}.

\bibliography{references.revtex}

\end{document}